\documentclass[preprint,10pt]{aastex}
\usepackage{psfig,dynamics}

\begin{document}

\title{On the Early Evolution of Forming Jovian Planets I: Initial
Conditions, Systematics and Qualitative Comparisons to Theory}
\author{Andrew F. Nelson\footnote{UKAFF Fellow. Present address: 
School of Mathematics, University of Edinburgh, Edinburgh Scotland
EH9~3JZ} }
\affil{Max Planck Institut f\"ur Astronomie, K\"onigstuhl 17, D-69117
Heidelberg, Germany}
\author{Willy Benz}
\affil{Physikalisches Institut, Universit\"at Bern, Sidlerstrasse 5, 
CH-3012 Bern, Switzerland}

\begin{abstract}

We analyze the formation and migration of an already formed
proto-Jovian companion embedded in a circumstellar disk. We use two
dimensional ($r,\theta$)\footnote{Throughout this paper and its
companion, \jtwo, we use `$\theta$' to denote the azimuth coordinate
rather than the more usual variable `$\phi$' in order to avoid
confusion between references to the coordinate and to components of
the planet's gravitational potential, $\phi_m$, common throughout
\jtwo.}\notetoeditor{It would be nice if the footnote numbering was not
reversed between the first footnote in the author address, and the
second, here in the abstract. Perhaps in a future AASTeX version?}
hydrodynamic simulations using a `Piecewise Parabolic Method' (PPM)
code to model the evolutionary period in which the companion makes its
transition from `\tone' migration to `\ttwo' migration. 

The results of our simulations show that spiral waves extending
several wavelengths inward and outward from the planet are generated
by the gravitational torque of the planet on the disk. Their effect on
the planet cause it to migrate inward towards the star, and their
effect on the disk cause it to form a deep (low surface density) gap
near the planet. We study the sensitivity of the planet's migration
rate to the planet's mass and to the disk's mass. Until a transition
to slower \ttwo\  migration, the migration rate of the planet is of
order 1~AU/10$^3$~yr, and varies by less than a factor of two with a
factor twenty change in planet mass, but depends near linearly on the
disk mass. Although the disk is stable to self gravitating disk
perturbations (Toomre $Q>5$ everywhere), implying the effects of
gravity should be insignificant, migration is faster by a factor of
two or more when disk self gravity is suppressed. Migration is equally
sensitive to the disk's mass distribution within 1--2 Hill radii of
the planet, as demonstrated by our simulations' sensitivity to the
planet's assumed gravitational softening parameter, and which also
crudely models the effect of the disk's extent into the third ($z$)
dimension.

Deep gaps form within $\sim500$~yr after the beginning of the
simulations, but migration can continue much longer: the formation of
a deep gap and the onset of \ttwo\  migration are not equivalent. The
gap is several AU in width and displays very nearly the \mplan$^{2/3}$
proportionality predicted by theory. Beginning from an initially
unperturbed 0.05\msun\  disk, planets of mass $M_{\rm pl}> 0.3$\mj\
can open a gap which is deep and wide enough to complete the
transition to slower \ttwo\  migration. Lower mass objects continue to
migrate rapidly for the duration of the simulation, eventually
impacting the inner boundary of our grid. This transition mass is much
larger than that predicted as the `Shiva mass' discussed in
\citet{PP4_WH}, making the survival of forming planets even more
precarious than they would predict.

\end{abstract}


\section{Introduction}

With the recent discovery of Jovian mass planets around other stars
\citep{51peg,47uma,70vir}, we find ourselves with a puzzle. To wit:
the two most reasonable scenarios (the `core accretion' and `collapse'
models) for forming Jovian mass planets fail when faced with the task
of forming a planet in short period orbits similar to those where many
planets are now being detected, because the gas temperature is very
high. Dust coagulation (the first step in the core accretion model)
cannot proceed because many grain species are in a gaseous state
rather than solid. Gravitational collapse cannot proceed because the
gas pressure is high and dominates over self gravity. It therefore
becomes attractive to form planets in one location, then somehow move
them to the locations where they are now being found.

Fortunately, mechanisms are known that are capable of causing the
migration of forming Jovian planets through the disk
\citep{LP79,GT79,GT80,HourWard84,LP86,PP3_LP,art93a,art93b,Takeuchi96,Ward97a}.
The problem is that these models actually work too well, predicting
migration of the planet through the disk and onto the star on only a
few thousand year time scale.

Recent research has focused on how to slow the migration so that the
planet can remain in orbit around its primary (without being accreted)
until after circumstellar disk dissipates. If it can survive this
long, the planet will likely be frozen into its final orbit,
presumably for the rest of the main sequence lifetime of the star.
\citet{LinBodRich96} for example, have examined the effect of tidal
resonances between the planet and the star within a few stellar radii
of the stellar surface. In another model, \citet{Trilling98} look at
the effects of mass transfer between the planet and star, in which a
planet can effectively trade part of its mass for increased orbital
distance from the star. Both find that for some conditions a planet
which migrates quickly to within a few stellar radii of the star can
indeed remain there without falling into the star for a time longer
than the lifetime of the disk.

These models are limited to stopping the migration quite near the
stellar surface. They follow the evolution of the system for
$\sim10^6$~yr in one spatial dimension and with the assumption that
the disk is initially relatively unperturbed by the existence of the
planet. Two-dimensional studies of the dynamical activity in the disk
produced by a planet larger than a few tenths of a Jovian mass
\citep[see e.g.][]{Kley99,Bryden99,AL96} show that it is inherently
multi-dimensional in nature and therefore not amenable to the one
dimensional analyses so far employed. Strong spiral structures
develop, which quickly redistribute mass and angular momentum and
under some conditions, cause a deep gap in the disk to form around
the planet. After gap formation, migration will slow considerably
because little mass is left near the planet with which to exchange
angular momentum. In many previous works however, the migration of the
planet through the disk and the effect of the migration on the
dynamics and system morphology were neglected in favor of studying
the gap formation process in isolation.

In this work, we will use a multi-dimensional hydrodynamic code to
simulate the evolution of an already formed planet migrating through
a circumstellar disk, concentrating in particular on the `\tone'
migration and gap formation epoch. Our overall strategy will be first
(section \ref{sec:model}) to outline initial conditions and the
physical model under which our simulations proceed, then (section
\ref{sec:phys-signif}) to define the systematic and physical
limitations that our simulations have. We will show that a number of
systematic errors may affect our results in important ways, but that
in most cases these errors are not fatal. Indeed, we will be able to
learn much from them if we properly account for their behavior and
physical significance in our simulations. In section \ref{sec:evol},
we describe the results of our simulations in qualitative detail and
the manners in which they do and do not agree with the predictions of
analytic and semi-analytic theories. In section \ref{sec:summary}, we
compare our results to others in the literature. A companion paper
\citep[][hereafter \jtwo]{jov2} will focus on a subset of the
simulations discussed here. It will examine the character of the mass
accretion and make detailed comparisons to analytic calculations of
the gravitational torques. 

\section{The physical model}\label{sec:model}

In this section we define the initial conditions at which the models
begin their evolution, the physical assumptions under which the models
evolve, the numerical methods used to carry out the evolution and the
limits they place on what we can learn. The initial conditions and
physical assumptions implemented in this work are quite similar to
those of the more massive disks discussed in \citet{DynamI}, so we
merely summarize the conditions specific to the current work here. 

\subsection{The initial conditions and equation of state}\label{sec:initconds}

We assume a self-gravitating, gaseous disk with mass \mdisk=.05\msun\
orbits a central star of mass $M_*=1.0$\msun. The gas in the disk is
set up in rotational equilibrium so that the radial accelerations due
to pressure gradients, gravity and centrifugal forces exactly cancel
each other. The surface density and temperature are power laws defined
as
\begin{equation}\label{eq:dlaw-mig}
\Sigma(r) = \Sigma_1 \left({{1 AU}\over{r}}\right)^p
\end{equation}
and
\begin{equation}\label{eq:tlaw-mig}
T(r) = T_1 \left({{1 AU}\over{r}}\right)^q
\end{equation}
where $\Sigma_1$ and $T_1$ are the surface density and temperature at
1~AU respectively. The surface density power law exponent is set to
$p=1.5$ and the temperature power law exponent is set to $q=0.5$.

The surface density coefficient, $\Sigma_1$, is determined from the
assumed disk mass and the radial dimensions of the disk. With the
given mass and dimensions (see section \ref{sec:the-code} below), we
obtain a surface density of $\sim$800~g/cm$^2$ at 5.2~AU, which is
similar to the value used for the total surface density in the
evolutionary models of \citet{Pol96}. In those models, Jovian planet
cores required several times $10^6$~yr to accrete enough mass in $Z$
element materials (i.e. non-H/He) to then accrete gas and form a giant
planet. We choose the temperature scale, $T_1=250$~K, to be typical of
the temperature profiles for young disks discussed in \citet{BSCG}.
With this choice, the derived isothermal scale height of the disk ,
$H=c_s/\Omega$, normalized by the distance from the star is
$H/r=0.045$.

\begin{figure}
\psfig{file=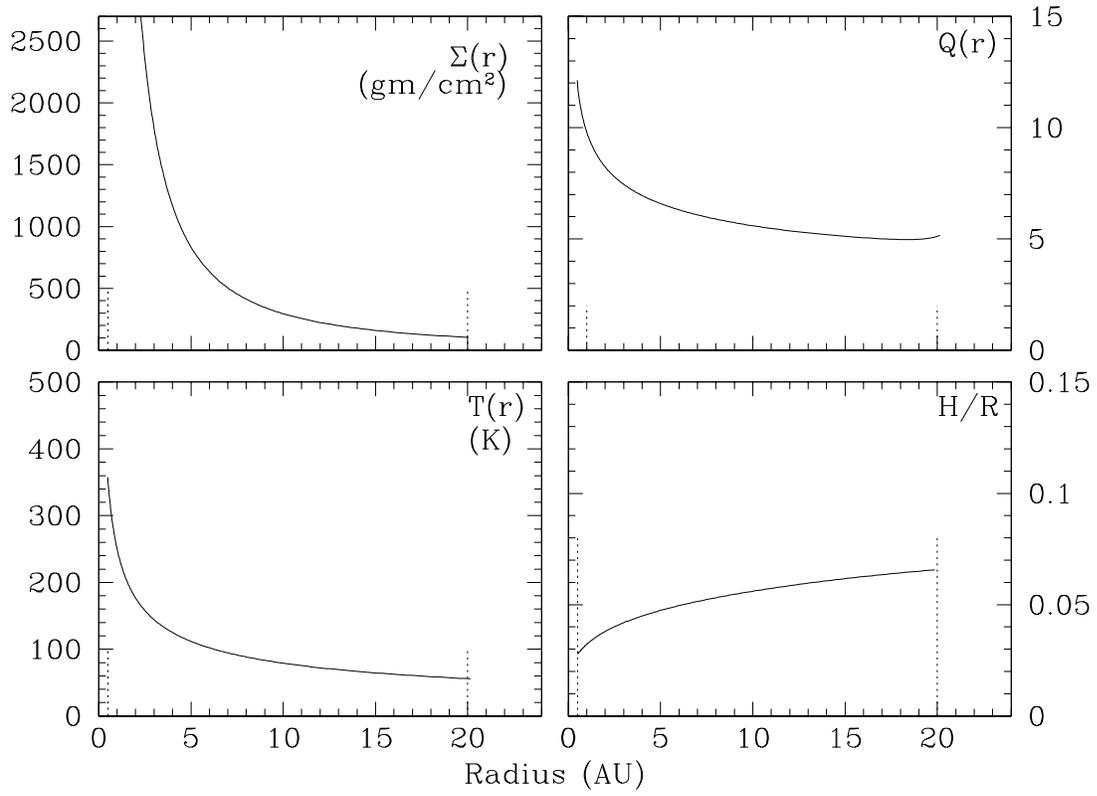,width=15cm,angle=-90}
\caption[Initial Conditions for the disk]
{\label{fig:disk-init}
Initial conditions for surface density, Toomre $Q$, temperature and
dimensionless scale height $H/r$ for the disk models. The dotted
vertical lines denote the inner and outer grid boundaries. }
\end{figure}

No initial perturbations are assumed in any of the models. A graphical
summary of the initial conditions for our simulations is shown in
figure \ref{fig:disk-init}. The upper right panel shows the Toomre
stability parameter, $Q$, value for the simulations at each radius.
Because it is $Q>5$ for all radii, we expect that although they are
included, the effects of self gravitating disk instabilities on the
evolution will be negligible. Note however, that this statement is not
equivalent to concluding that self gravity will have little effect on
the results.

As in \citet{DynamI}, we use a `locally isothermal' equation 
of state so that the vertically integrated pressure is defined by 
\begin{equation}
P = \Sigma c_s^2 \end{equation} 
where the sound speed, $c_s$, is given by $c_s=\sqrt{\gamma RT/\mu}$, 
$\mu$ is the mean molecular weight of the gas and $R$ is the gas 
constant. In our version of the PPM algorithm, PROMETHEUS, a truly
isothermal equation of state is not easy to implement, so instead we
set the adiabatic exponent $\gamma=1.01$, rather than exactly unity.

\subsection{The hydrodynamic code}\label{sec:the-code}

We have adapted a version of the PROMETHEUS hydrodynamic code
\citep{FMA89,FMA91} based on the PPM algorithm of \citet{ColWood84}
for the current study. PPM uses a high order interpolation to
reconstruct the flow variables at each zone interface, which then
specify the input values used to solve a Riemann shock tube problem
there. The high order reconstruction is modified in regions of steep
discontinuities to preserve sharp structures. Effectively in such
regions, it uses lower order approximation to the flow as input to the
Riemann problem. Physical dissipation in shock structures is included
explicitly in the solution of the Riemann problem so that no
artificial viscosity is required for stability of the code, and none
is included. 

The motion of the gas and the planet is resolved in radius and azimuth
$(r,\theta)$ with a logarithmically spaced grid (i.e. $\delta r = C
r\delta\theta$; $C\sim 1$ and where $\delta r$, $\delta \theta$ are the
dimensions of one zone). Two opposing conditions define the resolution
requirements for the system. First, we require that waves excited at
the Lindblad resonances of the planet be able to propagate through the
disk and dissipate before encountering a boundary and second we
require good spatial resolution of the flow in the neighborhood of the
planet.

We will find that waves are able to propagate inward from the planet
to $\sim1$~AU and outward to $\sim15$~AU, so in order to satisfy the
first condition we model the gas flow in the region between 0.5 and
20~AU. We implement reflecting boundary conditions at both the inner
and outer boundaries, so that any unphysical boundary interactions
will be made obvious by their large perturbations to the system.  

We monitor the gas near the boundaries for such large perturbations
and in general, we find that the inner boundary is more difficult to
handle than the outer. Our standard resolution simulations experience
little difficulty due to either boundary. In the high resolution
simulations however, waves are able to propagate all the way to the
inner boundary and be reflected. The amplitude of these reflected
waves is very small so that they do not directly affect the
migration. The main effect of the inner boundary on the simulation
will be to slow slightly the migration of a planet through the disk,
when and if disk matter cannot continue to accrete inward toward the
star. Instead it builds up between boundary and planet's orbit, and
produces larger positive torques on the planet that would otherwise
be the case. We terminate the simulations when such boundary effects
are judged to be seriously affecting the calculation. 

To satisfy the second condition we use a grid with 128 zones in radius
and 224 zones in azimuth as our standard resolution. With the inner
and outer boundary locations and our standard resolution, the linear
grid resolution near the planet (initially located at $r=5.2$~AU) is
$\sim0.12-0.15$~AU per zone. For our low and high mass prototype
simulations (0.3 and 1.0\mj\  planets respectively--see definition in
section \ref{sec:evol} below), this means that the Hill sphere of the
planet is resolved by about 4-5 or 7-8 zones respectively, depending
on exactly where the planet is in the disk. To test the effects of
resolution on our results we also run a number of simulations at
double this number of zones in each dimension. With our double
resolution simulations, the Hill sphere of a 0.3\mj\  planet is
resolved with about 15-20 zones.

We determine the gravitational force due to the matter in the disk on
itself using a two dimensional FFT technique for cylindrical grids as
described in \citet{GalDyn}. The forces between the point masses on
each other and between the point masses and the disk are described in
the following section.

\subsection{The point mass}\label{sec:pm-calc}

The planet is assumed to be orbiting the star at an initial orbital
radius of $a=$5.2~AU in an initially circular orbit. This yields an
initial orbit period like that of Jupiter, \tj$\approx$12~yr. In the
discussion below, we express the time units in terms of years or in
\tj. The planet's trajectory through the disk may be affected by
gravitational forces from the disk and the gravitational force due to
the central star. We do not include the effects of gas pressure forces
on the trajectory. The trajectory is computed at each time step using
a second order leapfrog method. 

We do not follow the star's trajectory but instead fix it to the
origin. In essence, by neglecting the star's motion means that we
neglect an indirect potential term due to the fact that the
star$+$planet system's center of mass is offset from the star's
center. This term is small in comparison to all other potential terms
everywhere in the disk except the planet's direct potential term in
regions where that term is itself neglectable. Therefore we expect
this approximation to cause little distortion of the results of
our simulations.

Fixing the star to the origin effectively assumes that the mass of the
rest of the system is small compared to the star, so that the system
center of mass and star are nearly coincident. In our own solar
system, the sun orbits the system center of mass at a distance of
about one solar radius. Thus by fixing the star to the origin we would
expect motion of the system center of mass due to neglect of the
stellar orbital motion (an error) of this magnitude in our
simulations. Additional contributions could come from inhomogeneities
in the disk, should they develop and interact with the grid
boundaries. We have monitored the system center of mass in our
simulations and found that it moves no further than 1--1.5\rsun\  from
the origin, depending strongly on the assumed planet mass, and in
accordance with our expectation. Thus, we believe that the sources of
error from neglecting the reflex motion of the star and from the
boundary interactions themselves are small.

The mutual forces exerted by the disk and the planet on each other are
calculated by direct summation of the contribution due to each grid
zone. We use a Plummer softened force law to determine the mutual
gravitational force, defined by
\begin{equation}\label{eq:pm-force}
{\bf F} =  -{{GM_{\rm pl} M_{\rm zone}}
           \over{r^2 + (\epsilon \delta)^2}} \hat {\bf r} 
\end{equation}
where $\hat {\bf r}$ is the unit vector joining the planet and the
center of a grid zone. The mass in each grid zone is $M_{\rm
zone}=\Sigma_{ij}r_i\delta r_i\delta\theta_j$, where $i$ and $j$ define
the index in the radial and azimuth directions respectively. The value
$\epsilon$ defines the softening radius of the potential, in units of
the local grid spacing, $\delta$. The size of one grid zone is
$\delta=\sqrt{(\delta r_i)^2 + (r_i\delta\theta_j)^2}$. The same formula
is used for the calculation of the force of each point mass on the
other, but in this case the softening radius is defined with a
constant value of 0.1~AU. 

In appendix \ref{app:pm-force}, we will find that the optimal value
for $\epsilon$ (from eq. \ref{eq:pm-force}) is at least 0.75 times the
physical size of a grid zone, $\delta$. In this case, the region
affected by the softened force law is limited to a region within a
distance of $\sim$2--3 grid zones from the planet. Unless noted
otherwise, we will choose a softening value of $\epsilon=1.0$ for our
standard resolution models and $\epsilon=2.0$ for our high resolution
models, so that we are well away from the numerically suspect (small
softening) regime. This value corresponds to a softening radius at
5~AU of $\sim0.15$~AU.

\section{What our models imply for physical systems}\label{sec:phys-signif}

Among the most important physical processes incorporated into our
models are the manner in which energy dissipation is handled and the
handling of the gravitational forces between the disk and planet. In
this section, we will first discuss the importance of dissipation in
the disk and how it is modeled in our code, then discuss the
sensitivity of the migration to the disk's small scale mass
distribution and how we can assess its importance. Finally we will
define the goals that we will be able to accomplish given our 
initial and physical conditions. 

\subsection{Implementation of dissipation in the
disk}\label{sec:diss-asmpt}

Physically speaking, there are two important effects of the
dissipative processes at work during the migration of a planet. First,
waves generated by the planet dissipate and transfer their energy and
momentum to the gas. Second, turbulent diffusion of gas transports
angular momentum and mass throughout the disk. In order to understand
the significance of the results of our simulations, it is important to
understand how both processes are incorporated into our models.

\subsubsection{Modeling dissipation due to shocks and
turbulence}\label{sec:diss-num}

As a consequence of their gravitational interactions with the disk,
protoplanets generate spiral waves. For planets more massive than
10--30\me\  (always the case in this work), interactions are strong
enough to drive some nonlinearity in the waves very close to the
planet so that shocks will develop there \citep{KP96,GR01}. The PPM
method has been designed to handle such shocks as well as possible. In
particular, the absence of any required artificial viscosity and the
presence of discontinuity sharpeners ensure that these features remain
sharp (2-3 zones) and propagate correctly. We therefore expect that
dissipation in shocks is handled well by our numerical approach.

In addition to dissipation in shocks, disks are believed to have an
internal dissipation mechanism that eventually lead to mass and
angular momentum redistribution even in absence of shocks. The exact
nature of the physical mechanism leading to this dissipation is a
matter of long debate (magnetic instability, convection, etc.). For
lack of a better model, it is customary to turn towards a so-called
"alpha-model" where the viscosity is simply parametrized by an unknown
factor $\alpha_{SS}$. No matter its exact origin or spatial variation,
several models \citep{BCKH,dAles,CG97,DynamII} have shown that a large
part of the radiated output of accretion disks is passively
reprocessed radiated stellar energy: internal dissipation in the disk
is small. Since the internal dissipation is small, the $\alpha_{SS}$
required to model it must also be small. In the context of our
migration models, this means the disk will be only weakly diffusive,
and that the expected effects from this dissipation source will be
small or equivalently affect the evolution only over long timescales.
On the other hand, the timespan simulated is necessarily short
compared to the lifetime of the disk hence we feel justified in
neglecting this type of dissipation.

Obviously, this does not mean that our code is viscosity free since
the transformation of the hydrodynamics conservation equations into
algebraic equations necessarily introduces unphysical dissipative and
dispersive terms. The key to meaningful simulations is that these
terms must remain small enough not to affect the evolution of the
system over the timescale simulated. In appendix \ref{app:ppm-diss},
we quantify the effects of the numerical dissipation and mixing
present in our code in the context of disk simulations including
spiral waves. We find that the numerical dissipation is very steeply
wavelength dependent and that dissipation of all but the longest waves
(lowest $m$ patterns) may indeed be quite large in our standard and
even in the high resolution simulations. Since PPM is a numerical
scheme with intrinsically low numerical viscosity, we expect that
effects of similar or larger importance will affect all numerical
simulations with equivalent or lower resolution.

In conclusion, in our numerical approach (as well as in most others)
waves are essentially dissipated locally. Correctly so for shock waves
and partially incorrectly so for low pressure waves which are affected
by numerical viscosity. However, in most analytical work local
dissipation is also assumed, waves are damped locally so that all
their energy and angular momentum is deposited at or near the
resonances. This together with the fact that our numerical results
depending most directly on dissipation (e.g. the gap widths discussed
in section \ref{sec:type1-2} below) are in good agreement with
analytical work indicate that local dissipation is what matters and
not the details on how this dissipation is actually taking place.

\subsubsection{Importance of the equation of state}\label{sec:diss-phys}

The physical assumptions which lead to dissipation in this work are
identical to those in \citet{DynamI}. Namely, that heating and cooling
are assumed to be locally isothermal. The temperature of any packet of
gas is defined by equation \ref{eq:tlaw-mig} as a function of distance
from the star and is fixed at the beginning of each simulation. The
physical consequences of a locally isothermal evolution are discussed
in detail in \citet{DynamII}. 

Briefly summarized, there is a delicate, {\it predefined} balance
between thermal energy input to the gas by dissipative heating (i.e.
shocks) and thermal energy lost to the gas by radiative cooling. This
equation of state is quite consistent with the idea of efficient
radiative damping of spiral waves suggested by \citet{CW96} and
reinforces the assumption of local damping discussed above. Changing
the absolute scale of the temperature power law or its exponent ($T_1$
and $q$ in eq. \ref{eq:tlaw-mig}), or assuming a `locally adiabatic'
equation of state instead as in several other works
\citep[e.g.][]{Boss98,Pick00}, will not modify the basic dissipation
assumption of the equation of state as long as we do not specifically
include heating and cooling based on physically relevant mechanisms.
Unfortunately, our limited understanding of these mechanisms makes a
more complex treatment of limited value and we limit ourselves only to
the simplest treatment here.

\subsection{Probing the disk's mass distribution close to the
planet}\label{sec:PM-signif}

Since from theoretical work we expect most of the important resonances
between the planet and disk to be located relatively close to it in
the radial coordinate, we naturally also expect the migration to be
very sensitive to the mass distribution in that region. The strongest
interactions will be at the Lindblad and corotation resonances, of
which the most important (i.e. the $m\sim10-20$ Lindblad resonances)
are found at about a disk scale height radially inward and outward of
the planet \citep{art93b}. For our initial conditions the scale height
is $H\sim0.23$~AU at 5~AU, and is quite comparable to the Hill radius
($R_{\rm H}\sim0.23$~AU or $\sim0.35$~AU for 0.3\mj\ and 1.0\mj\
planets, respectively). However, at the scale of $\sim1-2$\rh, the
assumptions underlying analytic theories (that the perturbation on the
disk is small) begin to break down. To what extent do interactions
from this region dominate the migration? Furthermore, on a $\sim1H$
distance scale, `close' to the planet must also be taken in a three
dimensional sense since by its definition matter in a real disk is
extended over a vertical extent comparable to the scale height.

Our simulations model only two dimensions, which means that they will
have an effectively amplified gravity because all of the disk matter
is located in the $z=0$ plane rather than at high altitudes more
distant from the planet. To allow us to study the sensitivity to the
3d mass distribution, we will follow the practice discussed in
\citet{Ward88} and \citet{KP93}. Each, using slightly different forms,
have used a gravitational softening parameter as a crude proxy for the
reduced gravitational force due to the vertical structure. Varying the
softening will thus provide a tunable probe of the dynamical
significance of mass in both the disk plane and its vertical extent,
over a region comparable in size to the softening radius. 

Simulations performed at different grid resolution provide a second
probe of the sensitivity. In this case, rather than probing the
influence of a given mass distribution, changing the grid resolution
probes changes in the mass distribution itself. Thus, we can conclude
that the migration rate is sensitive to the small scale mass
distribution in three dimensions if the migration rate depends
strongly on the magnitude of softening, keeping resolution fixed. We
can extend this conclusion to the small scale mass distribution in two
dimensions if the migration rate also depends on grid resolution,
keeping softening fixed.

\subsection{Relevance of Initial Conditions, and Goals of this
Work}\label{sec:IC-rel}

In section \ref{sec:disk-ev}, we will find that a planet of more than
a few tenths that of Jupiter's mass forms a gap in only a few hundred
years. Such rapid evolution naturally begs the question of what
relevance these initial conditions have, since the evolution away
from them is so rapid. In this paper and in \jtwo, our purpose will be
neither to develop a self consistent picture of the formation
processes of a planet nor to understand it's long term (`\ttwo')
migration through the disk. Rather, we will focus on two more limited
goals. 

First, we will study a large range of parameter space, in which we
model an early stage of evolution during which the disk is more
massive and the planet is not yet completely formed. In order to
survive, all planetary systems must pass through this stage
successfully. We will attempt to determine what physical effects are
most important in this range of parameter space, and determine what
types of systems can survive long enough to transition from the
`\tone' to the `\ttwo' evolutionary stages. In order to provide
meaningful comparisons, we will always begin from an identical initial
condition, representative of the earliest stages of the evolution. 

Second, we will make detailed comparisons to analytic models, in
order to constrain the regimes in which their predictions are
accurate and the manners in which they may fail. Such models are in
general based in linear theories where perturbations from a smooth
initial state are small, but in many cases are used in systems where
the planet's mass is already a substantial fraction of that of
Jupiter, so that the perturbations become large. It is for these
systems that multi-dimensional effects are most important and one
dimensional models become unable to properly model the evolution. In
order to make such comparisons we will require initial conditions
similar to those assumed in their analyses.

\section{Evolution of the system}\label{sec:evol}

Using the initial conditions and physical model above we have
simulated several series' of systems. Two series vary physical
parameters of the planet mass and disk mass, two series vary the
numerical parameters of grid resolution and gravitational softening
and one series varies the physics itself, including or excluding disk
self gravity. The parameters for all of the simulations are summarized
in Table \ref{tab:table-mig}.

In this section we will describe the characteristics of a selection of
our simulations, typical of the results obtained for the other similar
runs. The first of these models we refer to as our `high mass
prototype'. It describes the evolution of a relatively large mass
planet through the disk (simulation {\it mas7}, with \mplan=1.0\mj,
at our `standard' grid resolution of $r\times\theta=128\times224$). We
will find that the evolution of this system away from its initial
state is very rapid and, following our initial discussion of it in
sections \ref{sec:disk-ev} and \ref{sec:migration}, we will limit
nearly all of our further discussion to simulations with lower mass
planets. 

The second model will be referred to as our `low mass prototype'
model, and will describe the evolution of a system with a planet with
\mplan=0.3\mj. Our results will show that including or excluding disk
self gravity has a large influence on the eventual fate of the planet,
and we will discuss two simulations which are identical except for
this effect. The first includes disk self gravity (simulation {\it
mas3}) and the second omits it (simulation {\it nosg}). 

The third model will be referred to as our `high resolution prototype'
model and is a high resolution ($256\times448$) counterpart of the low
mass prototype, again with (simulation {\it Sof7}) and without
(simulation {\it Nosg}) disk self gravity included. Unless
specifically noted, our discussion will always refer to the versions
with self gravity. 

\subsection{Evolution of the disk, under the planet's
influence}\label{sec:disk-ev}

\begin{figure}
\psfig{file=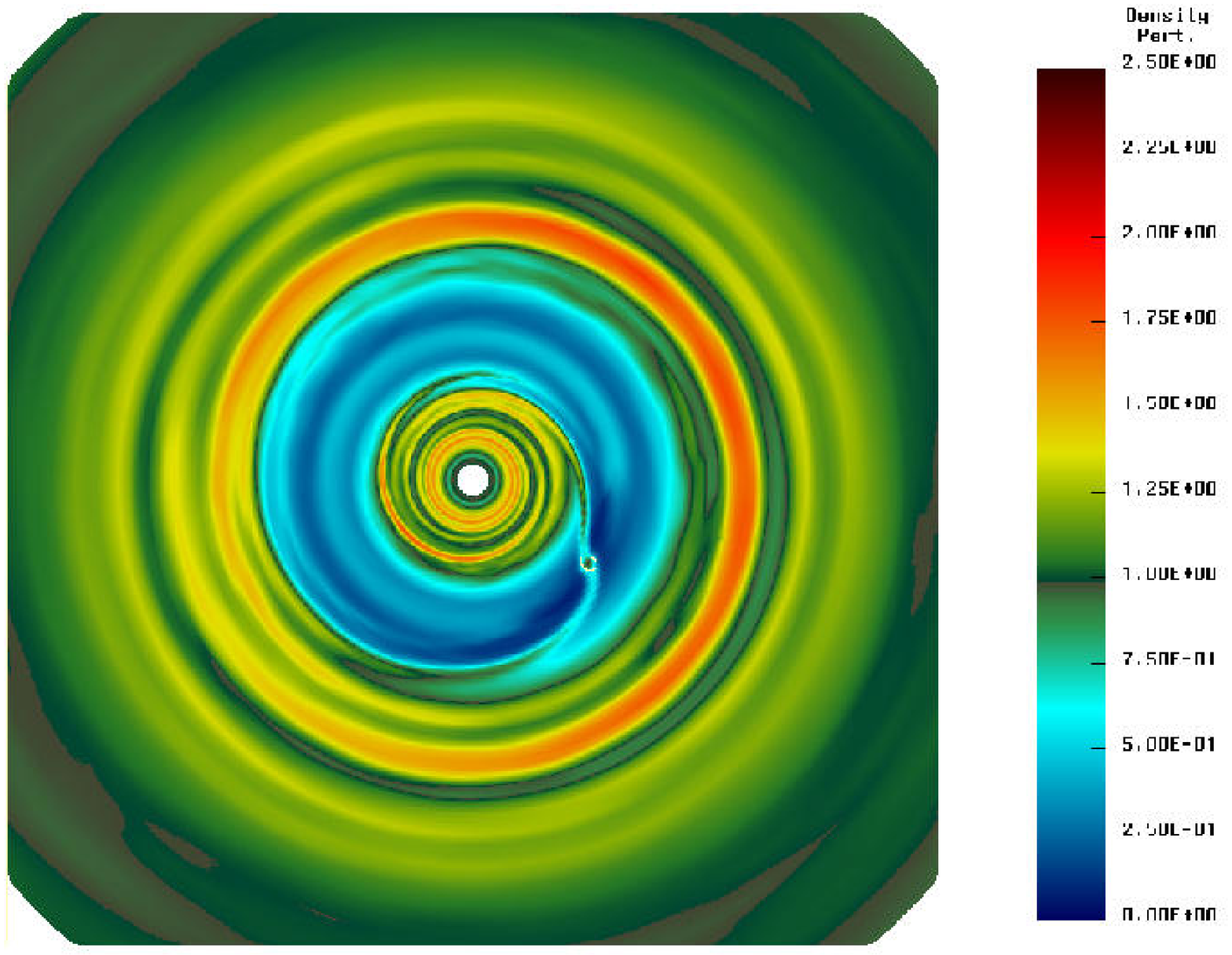,height=9.3cm,rheight=9.0cm,angle=-90}
\psfig{file=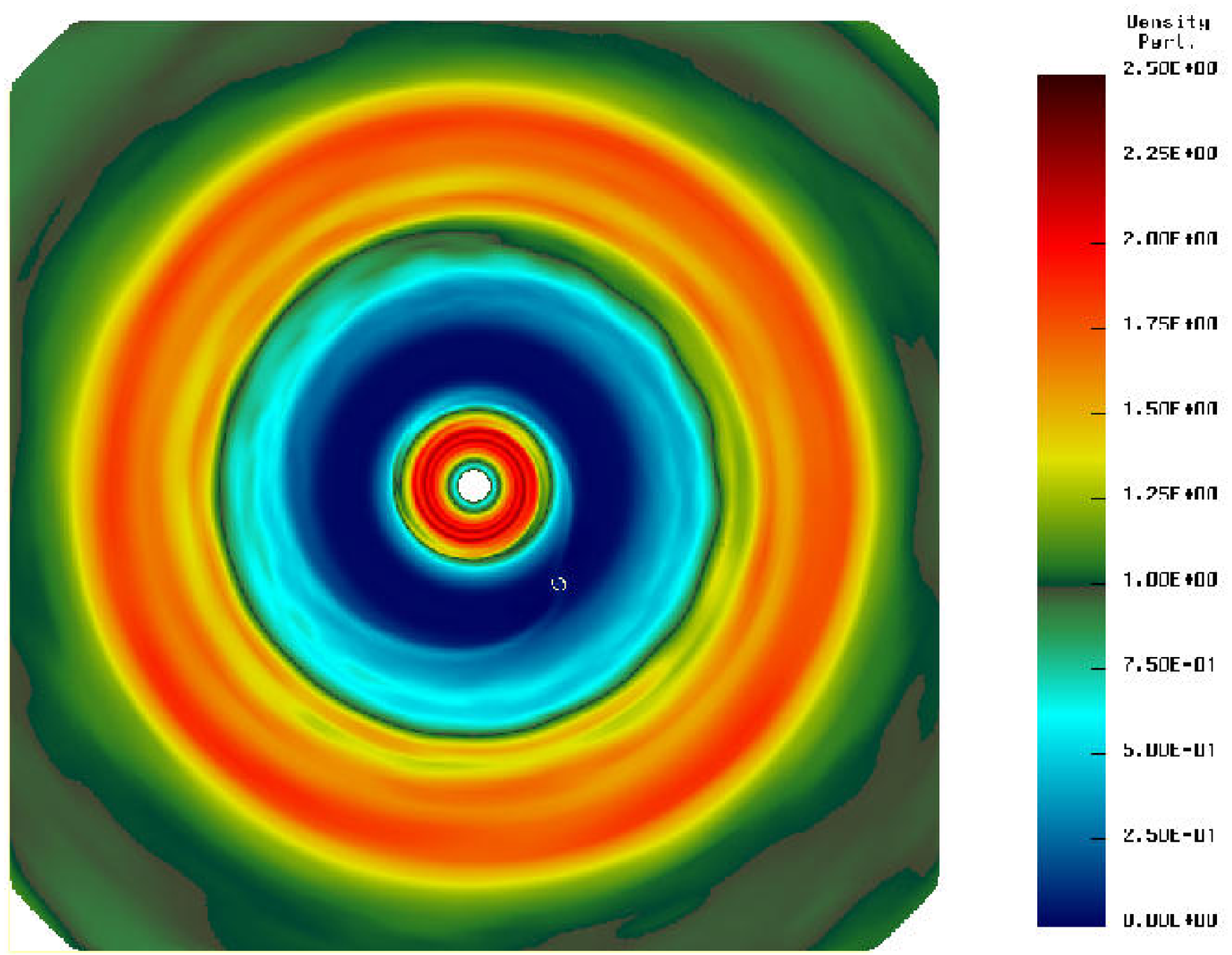,height=9.3cm,rheight=9.0cm,angle=-90}
\caption[Early and late snapshots of the disk density perturbations
caused by a 1$M_J$ planet traveling within the disk]
{\label{fig:himass-morph}
Perturbations of the surface density structure of the disk relative to
the initial density at each point
($\Sigma(r,\theta,t)/\Sigma(r,\theta,t=0)$)for the high mass
prototype simulation, {\it mas7}, in which an embedded 1\mj\  planet
sweeps out disk matter and begins forming a gap. (top) Early in the
run, prior to achieving a steady state (300~yr after the beginning of
the run). (bottom) Late time, near steady state density structure
(1800 yr). In both images, the boxed region $\pm$15~AU from the origin
is shown. The white ovoid shape defines the region dominated by the
gravity of the planet (i.e. its inner and outer extent define a
distance of 1\rh\ from the planet).} 
\end{figure}

\begin{figure}
\psfig{file=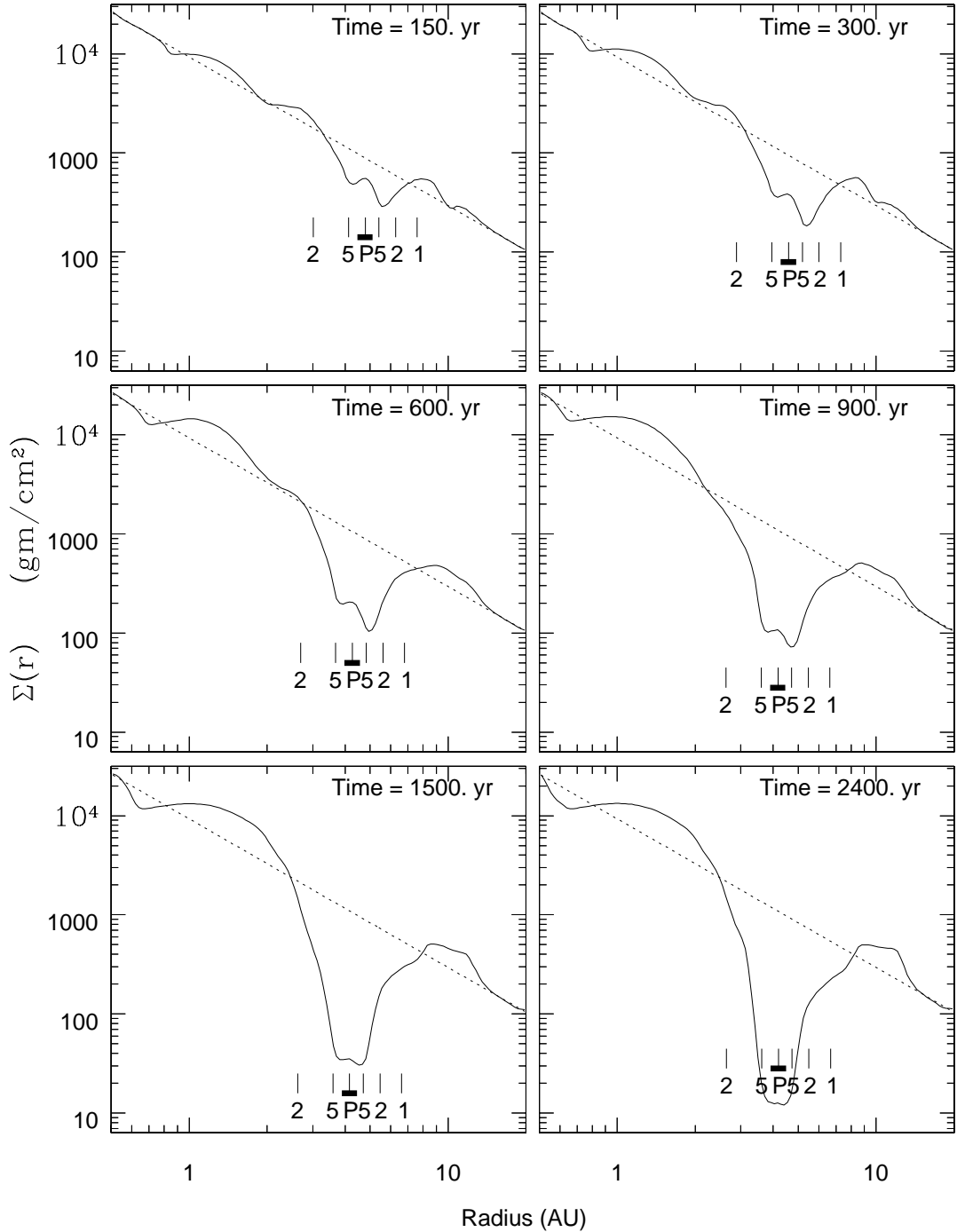,height=20.0cm,rheight=19.1cm}
\caption[Azimuth averaged disk density structure at several times]
{\label{fig:az-ave-dens-hi}
The azimuth averaged density structure of the disk at several times 
lines). The dotted line shows the initial density profile of the disk.
Also shown at each time are the location of the planet (P) and three of
the lowest order ($m=1,2$ and 5) Lindblad resonances induced by the
planet. The thick horizontal bar on the planet's symbol shows the extent
of Hill sphere of the planet.}
\end{figure}

Snapshots of our high mass prototype model are shown for early and
late times of the evolution in figure \ref{fig:himass-morph}. 
Figure \ref{fig:az-ave-dens-hi} shows the azimuth averaged surface
density profile at various times during the evolution. Early in the
simulation, spiral density waves are excited by the planet which
extend inwards from the planet's position to $\sim 1-1.5$~AU and
outwards to $\sim 15$~AU before fully dissipating. Within the first
$\sim100$~yr, a large evacuated region trailing the planet and
extending nearly a third of the way around the disk has formed in the
planet's wake. A similar region extends ahead of and radially inward
from the planet.

After another 2-300~yr of evolution, the evacuated regions ahead of
and behind the planet have grown completely around the star and have
begun to form a gap region in the disk where the surface density has
decreased a factor of 3--5 below its initial value. During its
formation, the gap is deepest inward/ahead of the planet and
outward/behind the planet in the sense of the orbital trajectory and
is not axisymmetric. Strong spiral structure is present near the
planet and throughout the disk. Finally after 6-800~yr of evolution,
the spiral structures decrease in amplitude, the disk becomes nearly
axisymmetric and settles into a slower, secular evolution. Vestiges of
low amplitude spiral structure are still present in the gap even after
1800~yr of evolution and extend inwards and outwards several AU 
into the higher density regions of the disk.

The gap forms quickly after the beginning of the simulation and within
600 years has grown to a 3~AU width, while the planet has migrated
inwards to 4~AU from the star. When the migration stops, the density
near the planet is a factor of six to eight below the initial density
at 4~AU, at about 200~g/cm$^2$ near the planet and 100 g/cm$^2$ near
the minimum, which is not found at the same orbit radius as the
planet, but slightly further out. Later, when the gap has more fully
formed, a very low density central region in the gap forms within
$\pm2R_H$ from the planet. Once the gap has formed, little further
evolution of the disk as a whole occurs, but the surface density near
the planet continues to decrease and falls to $<10$g/cm$^2$ at
$t=2400$~yr. The matter in the outer disk is left behind by the
migration of the planet inward, and the planet's influence on it
becomes progressively smaller. The gap's outer edge is thus much more
shallow than its inner edge, to which the planet moves progressively
closer until its migration stops. 

\begin{figure}
\psfig{file=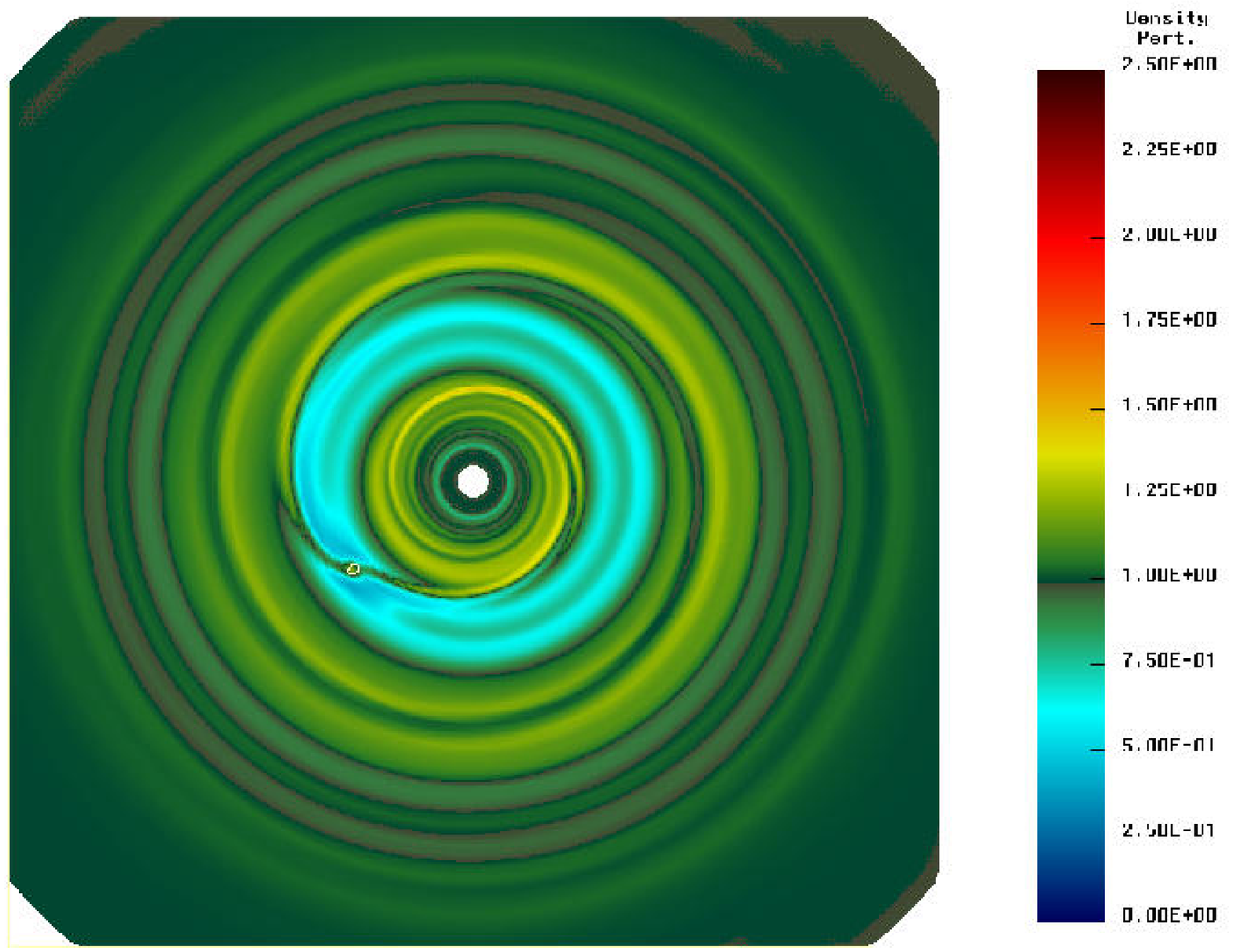,height=9.3cm,rheight=9.0cm,angle=-90}
\psfig{file=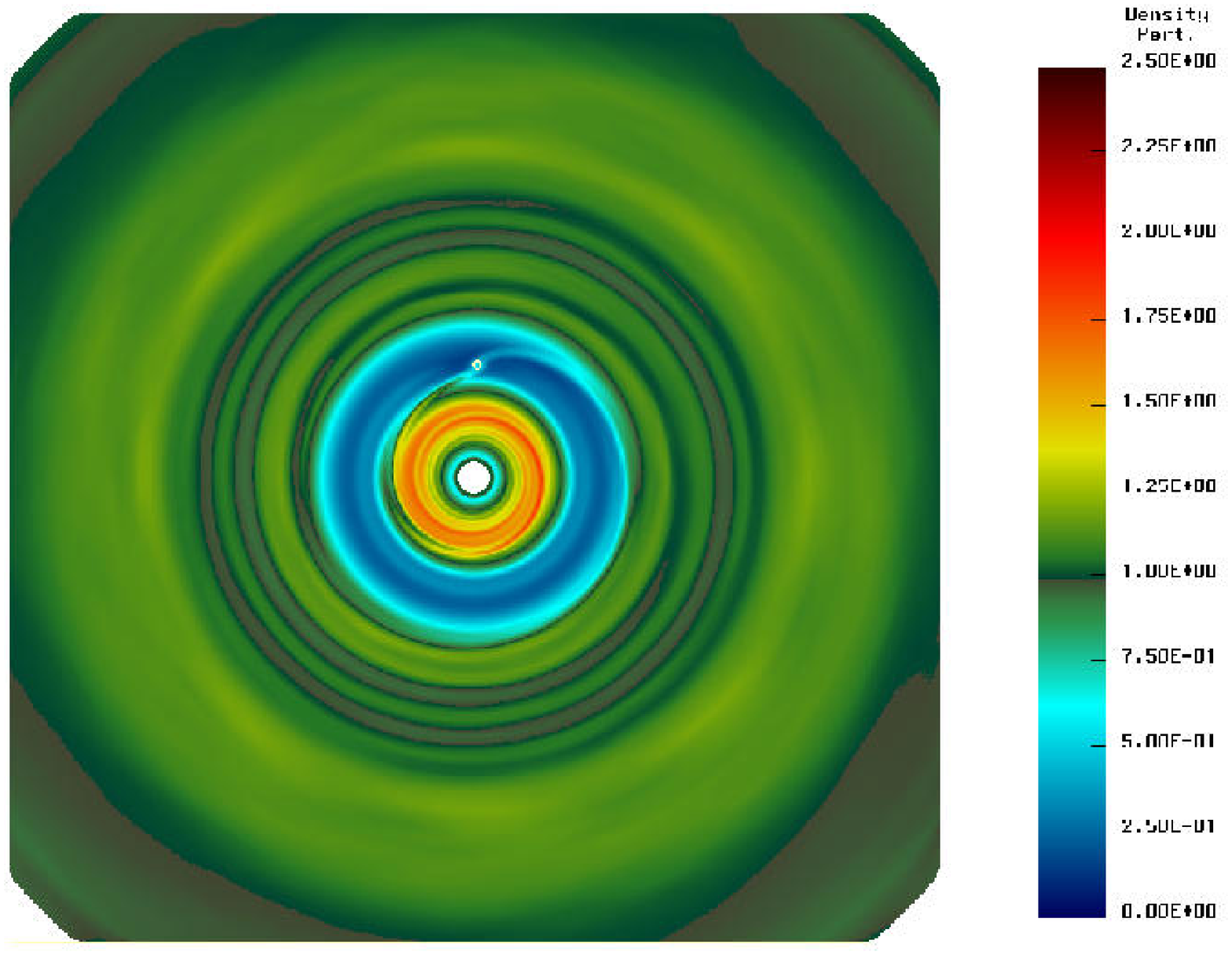,height=9.3cm,rheight=9.0cm,angle=-90}
\caption[Early and late snapshots of the disk density perturbations
caused by a 0.3\mj\  planet traveling within the disk]
{\label{fig:lomass-morph}
Perturbations on the surface density structure of the disk for the low
mass prototype (simulation {\it mas3}) after being perturbed by a low
mass planet (0.3\mj) traveling through the disk. The times from the
beginning of the run are 300~yr (top) and 1800~yr (bottom). In
comparison to the high mass prototype, spiral structures and
perturbations are much lower in amplitude. As in figure
\ref{fig:himass-morph}, the white ovoid shape defines the region
dominated by the gravity of the planet (i.e. its inner and outer
extent define a distance of 1\rh\  from the planet).} 
\end{figure}

The evolution of the low mass prototype is shown in figure
\ref{fig:lomass-morph}. As for the high mass prototype, the planet
builds spiral structure ahead of (behind) its azimuthal position in
the disk and which is inward (outward) from its radial position. The
spiral structure is of much lower amplitude in this case but remains
present for the duration of the simulation. It extends radially
inward from the planet to 1.5~AU and outward to 15~AU. In this model,
as for the high mass prototype, secondary, $m=2$ and $m=3$ symmetry
patterns are easily visible in addition to the primary spiral pattern
which intersects the planet's position in the disk.

\begin{figure}
\psfig{file=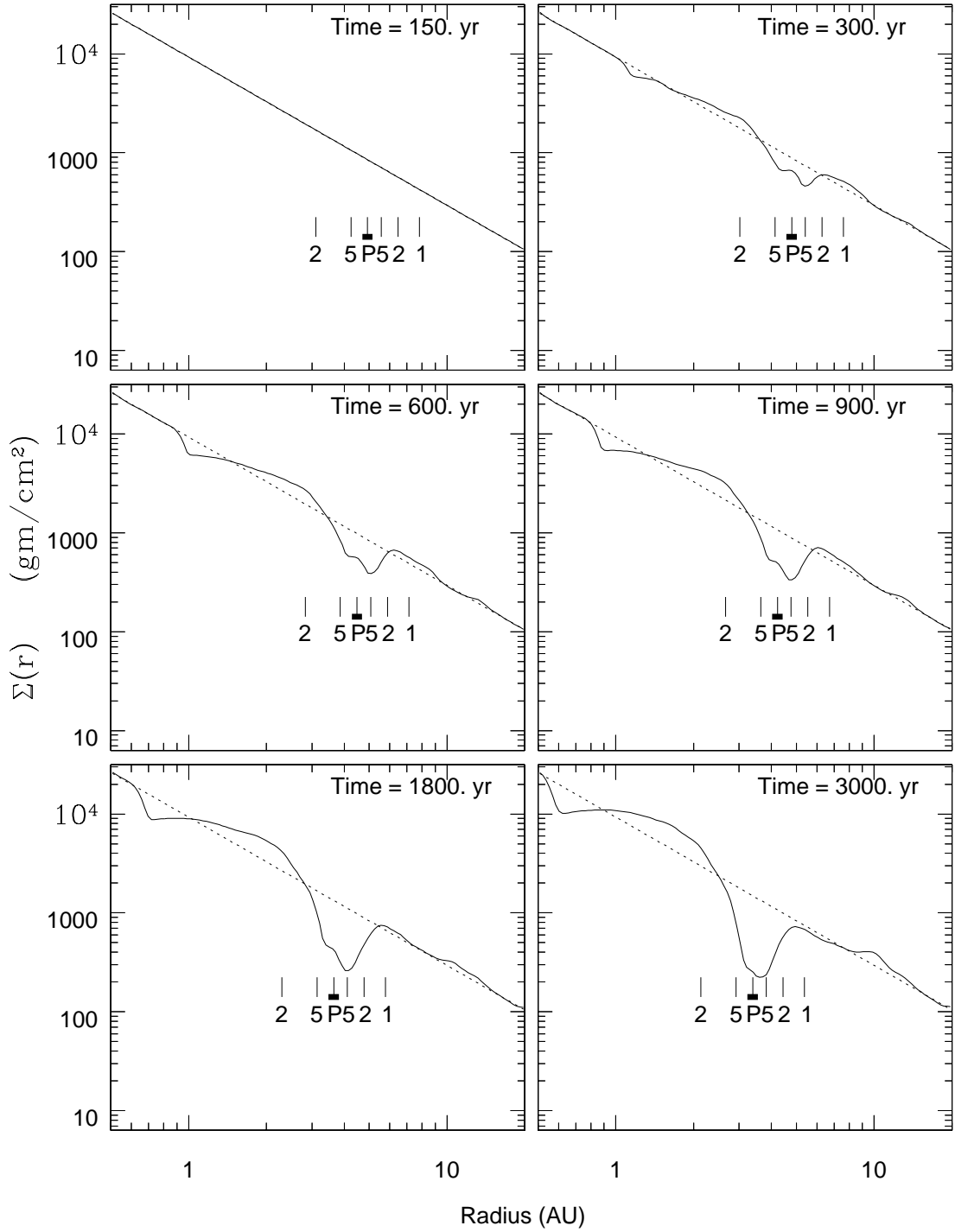,height=20.5cm,rheight=20cm}
\caption[Azimuth averaged disk density structure at several times]
{\label{fig:az-ave-dens-lo}
The azimuth averaged density structure of the low mass prototype
simulation at several times (solid lines). The dotted line shows the
initial density profile of the disk. Symbols are as in figure
\ref{fig:az-ave-dens-hi}.}
\end{figure}

Gap formation occurs more slowly in this simulation and the gap does
not reach the depth necessary to stop the planet's motion (see section
\ref{sec:migration}) during the first $\sim$2000~yr evolution, but
after 3000~yr of evolution, it does build a deep enough gap and rapid
migration stops. Over the course of the run the planet migrates inward
about 2~AU, and when the migration slows, the surface density in the
gap has decreased to a factor $\sim4-5$ below the initial density
(figure \ref{fig:az-ave-dens-lo}). The outer disk density profile
remains relatively unperturbed during the evolution.

\begin{figure}
\psfig{file=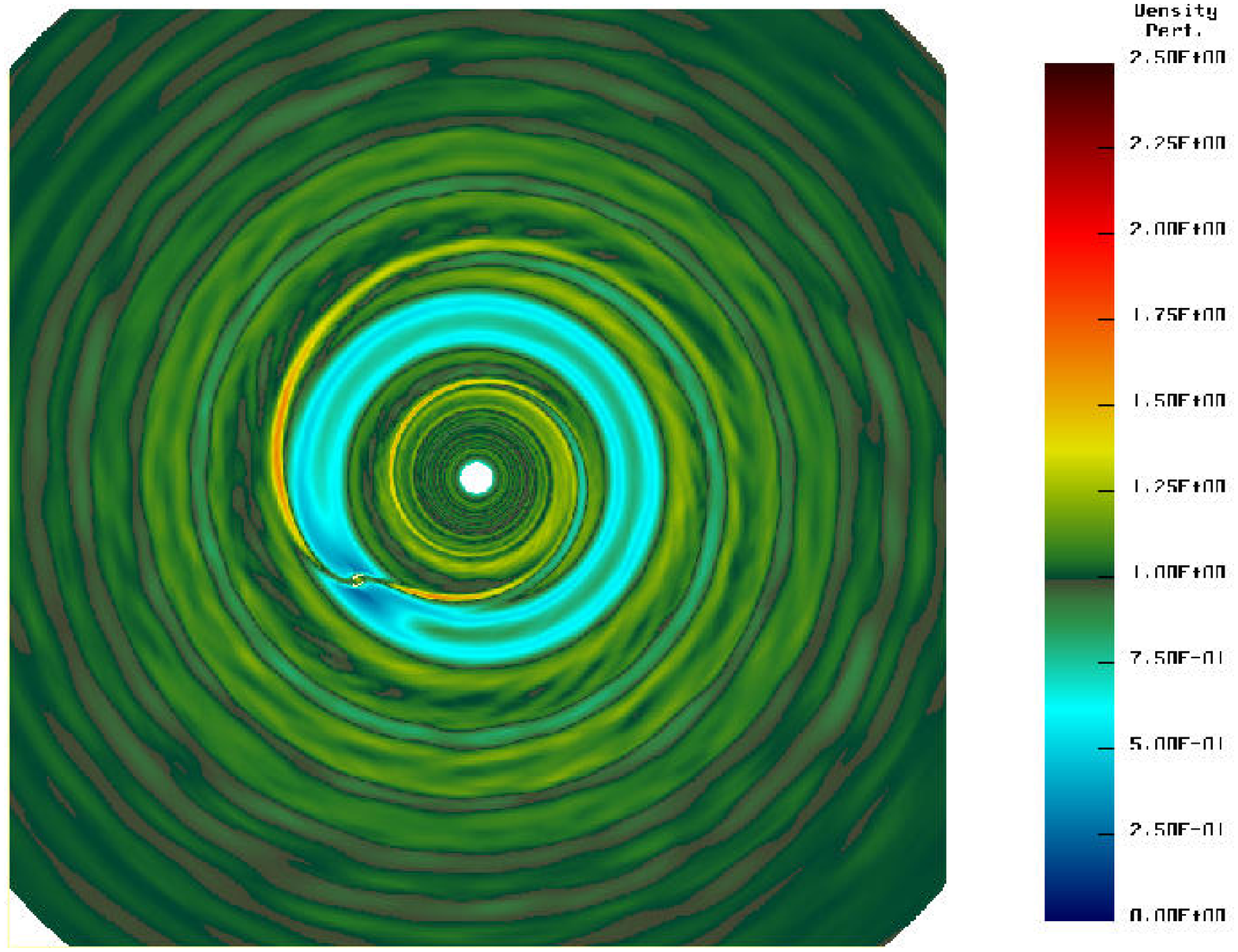,height=9.3cm,rheight=9.0cm,angle=-90}
\psfig{file=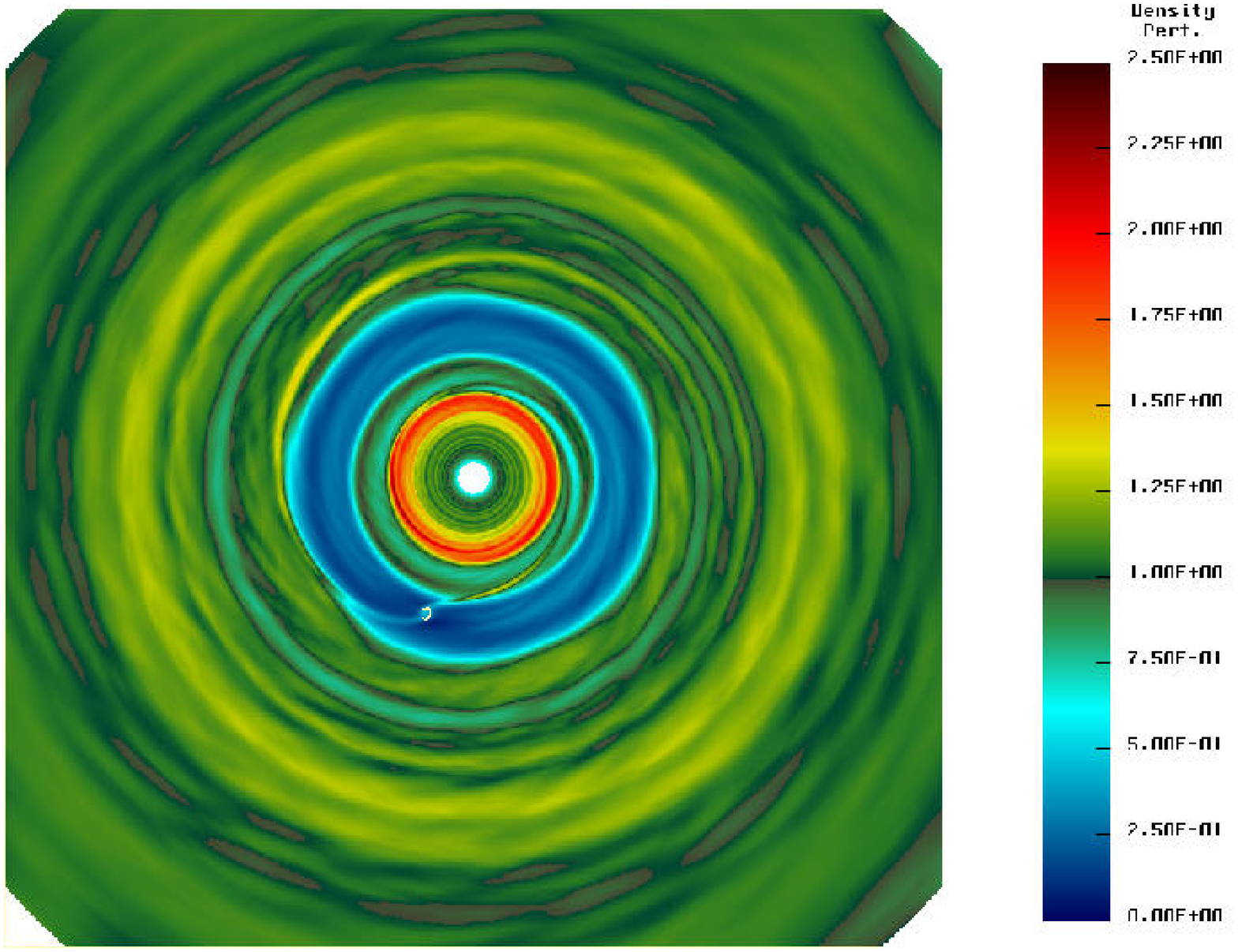,height=9.3cm,rheight=9.0cm,angle=-90}
\caption[Early and late snapshots of the disk density perturbations
caused by a 0.3\mj\  planet traveling within the disk]
{\label{fig:hires-morph}
Perturbations on the surface density structure of the disk for the
high resolution prototype (simulation {\it Sof7}) after being
perturbed by a low mass planet (0.3\mj) traveling through the disk.
(top) Early in the run, after 300~yr of evolution. (bottom) Late in the
run (1800~yr). In comparison to the low mass prototype, spiral wave
structures are visible much further from the planet and the gap is
slightly deeper and wider.  As in figure \ref{fig:himass-morph}, the
white ovoid shape defines the region dominated by the gravity of the
planet (i.e. its inner and outer extent define a distance of 1\rh\
from the planet). } 
\end{figure}

\begin{figure}
\psfig{file=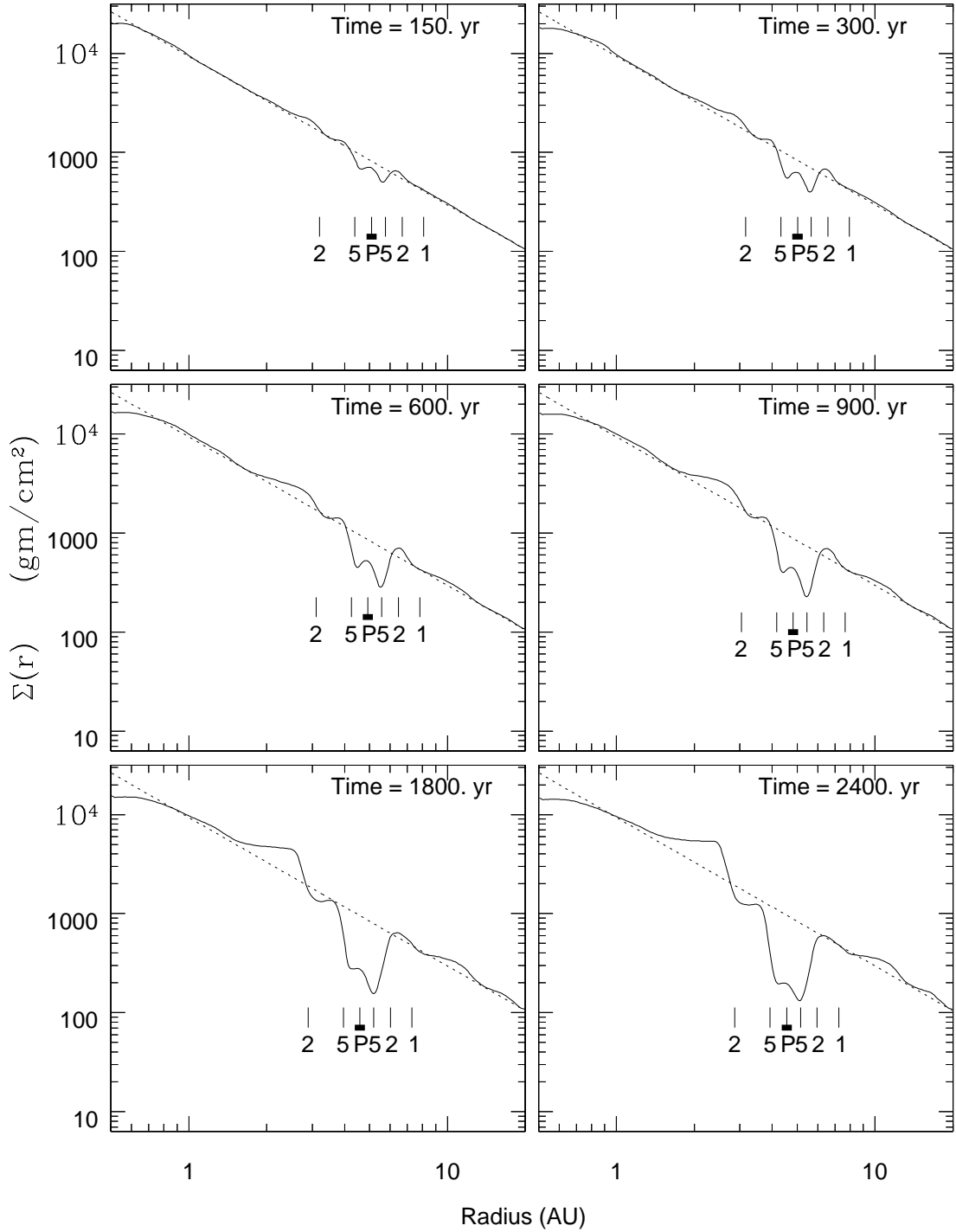,height=20.5cm,rheight=20cm}
\caption[Azimuth averaged disk density structure at several times]
{\label{fig:az-ave-dens-hires}
The azimuth averaged density structure of the high resolution
prototype simulation at several times during the evolution (solid
lines). The dotted line shows the initial density profile of the disk,
and symbols are as in figure \ref{fig:az-ave-dens-hi}}
\end{figure}

In order to make direct comparisons to our standard resolution runs,
we ran a high resolution simulation with the gravitational softening
parameter increased to $\epsilon=2.0$ (making it identical in
absolute size to the low mass prototype). We show early and late time
snapshots of the evolution of a run like our low mass prototype in
figure \ref{fig:hires-morph} and azimuth averaged surface density
profiles in figure \ref{fig:az-ave-dens-hires}. Qualitatively, many of
the features in figures \ref{fig:lomass-morph} and
\ref{fig:hires-morph} are quite similar in the region close to the
planet. In both examples, spiral waves excited by the planet are
visible inward and outward from the planet's position, a gap forms
around the planet and spiral structures extend into this gap region to
the planet. 

There are also important differences between the two simulations. In
the high resolution prototype, density perturbations are
nonaxisymmetric until much later times in the evolution and small
amplitude, high $m$ spiral structures are visible in the horseshoe
region of the gap. Spiral structure extends much further inwards and
outwards than with the standard resolution model, over nearly the
entire radius range simulated. Although the spiral structure extends
further, the changes that develop in the surface density profile are
more concentrated into distinct regions. The gap becomes deeper more
quickly and the region just inward of the $m=2$ inner Lindblad
resonance develops a large density enhancement as matter piles up
there. The differences that exist are due primarily to changes in the
breakdown point (i.e. which $m$ patterns propagate some distance,
rather than damp immediately upon formation) of our local damping
assumption and few physical conclusions can be made about them. 

At higher resolution, some wave components excited by the planet are
able to reach both the inner and outer grid boundaries with a small
but non-zero amplitude. Except to produce a perturbation on the
surface density profile there (figure \ref{fig:az-ave-dens-hires}), we
do not believe this boundary interaction strongly affects the gap
formation process or the planet's migration through the disk. Since the
waves are already very low amplitude at the boundary, their reflections
do not travel far before being completely dissipated. 

Perturbations of similar magnitude, clearly unrelated to boundary
interactions, are present in both the high and low mass prototype
simulations. In these runs, the inner disk develops a much flatter
radial profile than it initially has. The flattening in the profile is
due in part to the migration of mass inwards which was originally in
the gap region, but is also due to the outward migration of some
material in the inner disk. The location at which this is most
apparent ($\sim1.0-1.5$~AU--see the early time panels of figures
\ref{fig:az-ave-dens-hi} and \ref{fig:az-ave-dens-lo}) corresponds to
that at which the inner spiral structure decays to zero amplitude (see
\jtwo).

We were not able to isolate the exact physical or numerical origin of
the locally outward migration. However, we believe it is no
coincidence that the behavior is most visible at the same location as
the decay to zero amplitude of the spiral structure. While the largest
fraction of gravitational torque interaction is generated by the
planet near its Lindblad resonances, a small, but non-zero torque
contribution is generated far from the resonance. This contribution
oscillates in sign, successively adding or subtracting angular
momentum from the wave as it propagates away from the planet. Wave
decay implies a large asymmetry between crests and troughs of
successive radial waves and the torques they are responsible for, so
that the portion of this locally generated torque realized is not
symmetric from crest to trough of one wave form to the next, perhaps
leading to the observed effect.

\subsection{Evolution of the planet's orbit under the disk's 
influence}\label{sec:migration}

Because of the efficient transfer of angular momentum between the
planet and the disk, a massive planet experiences large dynamical
torques and suffers rapid inward migration. In some cases, its orbit
radius decreases by a third after only a few hundred years of
evolution. In figure \ref{fig:orb-migrate}, we show the radial
trajectory of the planet from our three prototype simulations
discussed above. After a $\sim100$~yr period in which the planet first
builds spiral structures, the high mass prototype (top panel) produces
a near constant migration rate for the next 500~yr. After 600~yr, when
the planet has reached an orbit radius of about 3.6~AU, the migration
rate decreases to near zero. Finally, the trajectory changes to a slow
outward migration as the planet's position relative to the gap center
equilibrate.

For the low mass prototype, the migration proceeds for most of the run
at a slowly decelerating rate. The migration begins to slow more
strongly after 2000~yr and, after 2700~yr, stops when a deep enough gap
forms. The version of this simulation without disk self gravity does
not decelerate and stop. Instead it migrates at a much faster rate so
that within 1000~yr has migrated more than 2~AU from its initial orbit
radius. Shortly thereafter (not shown on the plot), it begins to
accelerate as it moves into higher surface density regions closer to
the star and after 1200~yr, it impacts the inner boundary of our grid.
Similar behavior is observed in the simulations with very low mass
planets including disk self gravity (e.g. simulation {\it mas1} with
\mplan=0.1\mj).

In contrast, the high resolution prototype (with disk self gravity) is
characterized by a much slower migration rate than its standard
resolution counterpart. Over the entire duration of the simulation the
planet migrates only about 0.6~AU, compared to the $\sim2$~AU
migration that the standard resolution simulation suffers. The
migration also slows and stops due to the differences in the character
of the gap that forms. On the other hand, the version without disk
self gravity suffers an extremely rapid migration rate--faster than
either its low resolution counterpart or either of the two simulations
including self gravity. After only $\sim$300~yr, it has migrated 2~AU
through the disk and very quickly thereafter accelerates still faster
and begins to interact with the inner grid boundary.

In none of the four simulations with 0.3\mj\  planets, are migration
rates obtained that are identical to any of the others. Among the four
models, we vary two parameters: the presence/absence of disk self
gravity and sensitivity to the 2d mass distribution (via the varying
grid resolution). Looking at the two pairs of models with and without
self gravity at the same resolution (i.e. the pairs {\it mas3/nosg}
and {Sof7/Nosg}), we observe that in both cases the migration is
faster without self gravity than with it. This is interesting because
any variation at all runs counter to the expectation that self
gravitating disturbances should be largely suppressed, since the
Toomre stability parameter is $Q>5$ everywhere. On the other hand,
looking at the two pairs of models that vary grid resolution with or
without self gravity (i.e. the pairs {\it mas3/Sof7} and {\it
nosg/Nosg}), we do not observe a consistent pattern: with self gravity
migration is slower with higher resolution but without self gravity,
migration is faster with higher resolution.

It is significant that the rate differences in the latter case cannot
be due to changes in the dissipation between the models. Such
differences would produce large changes in the surface density
profile, as shown in figures \ref{fig:az-ave-dens-lo} and
\ref{fig:az-ave-dens-hires}. However, during the period the migration
rates are calculated, the gap has only begun to form. Its structure is
thus relatively undifferentiated from the others and will have little
effect on the torques on the planet. Therefore we can conclude that
the origin of the rate differences lies in physically meaningful
quantities: in one case the effect self gravity and in the other the
mass distribution near the planet.

\subsection{Dependence of the migration rate on planet mass and
disk mass}\label{sec:rates}

We expect the migration of a planet through the disk to depend both on
the mass of the planet and the mass of the disk in which the planet
evolves. The \tone\ migration rate (expressed as a migration velocity
by \citet{Ward97a}) is directly proportional to the planet mass and
the disk surface density 
\begin{equation}\label{eq:mig-timescale}
{{\dot a}\over{a}} \propto M_{\rm pl}\Sigma,
\end{equation} 
given that only Lindblad resonances are important for the migration.
We have run a series of simulations varying planet mass (designated
{\it mas} in table \ref{tab:table-mig}) in order to test this
proportionality. Further, to test the sensitivity of the migration
rate to the mass surface density, we have run a series of simulations
(designated {\it dis}) which vary the assumed disk mass. With
identical grid dimensions (inner and outer boundary radii), disk mass
and surface density are directly proportional, so that the
proportionality in equation \ref{eq:mig-timescale} still holds. 

\begin{figure}
\psfig{file=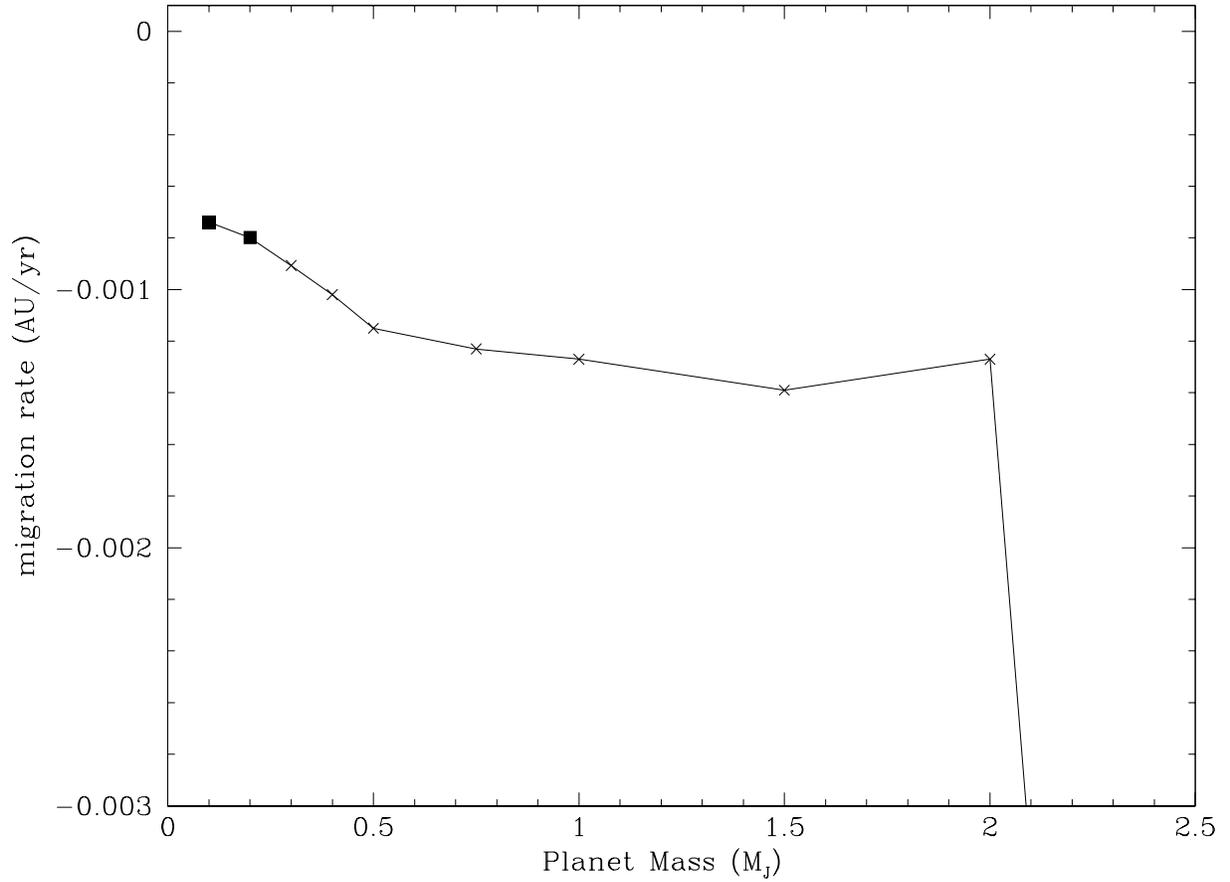,height=5.0in,angle=-90}
\caption[Planet mass dependence of migration rates]
{\label{fig:rates-mass}
The migration rate of planets with various masses. The planets for
which a gap was opened and the migration rate decreased to near zero
are shown with a cross, while planets that did not open a gap are
shown with a filled square. Planets with masses larger than 2.0\mj\
were affected by numerical errors (see text) and migrated so rapidly
that no gap could form and the planets fell into the star.}
\end{figure}

In figure \ref{fig:rates-mass}, we show the migration rates fit for
simulations with different planet masses. The rates shown are valid
for the period for which the migration rate is constant in that
simulation (see figure \ref{fig:orb-migrate}). In other words, the fit
is obtained over the period for which the migration can be considered
`\tone', and the planet has migrated less than $\sim$2~AU from its
original orbit radius. The rates are nearly independent of planet
mass--the expected linear trend is not present. The migration rate for
the simulation with \mplan$=4$\mj\ is plotted to complete the full
series, but should be disregarded since it was affected by a
numerical error (see appendix \ref{app:pm-force}).

The only significant feature in figure \ref{fig:rates-mass} for
planets below 2\mj\  is a small `knee' in the slope and flattening in
the rates above 0.5\mj. This feature may be explained by a close
examination of figures \ref{fig:himass-morph}, \ref{fig:lomass-morph}
and \ref{fig:hires-morph}, which shows that the perturbation amplitudes
have reached 20\% or more, so that for these masses the planet is
essentially driving the waves to saturation so that the linear
theories no longer apply. Furthermore, at such high planet masses, the
gap formation occurs so quickly that a truly \tone\ migration epoch,
with an unperturbed disk, may not exist for a period long enough to
obtain a well defined migration rate. 

\begin{figure}
\psfig{file=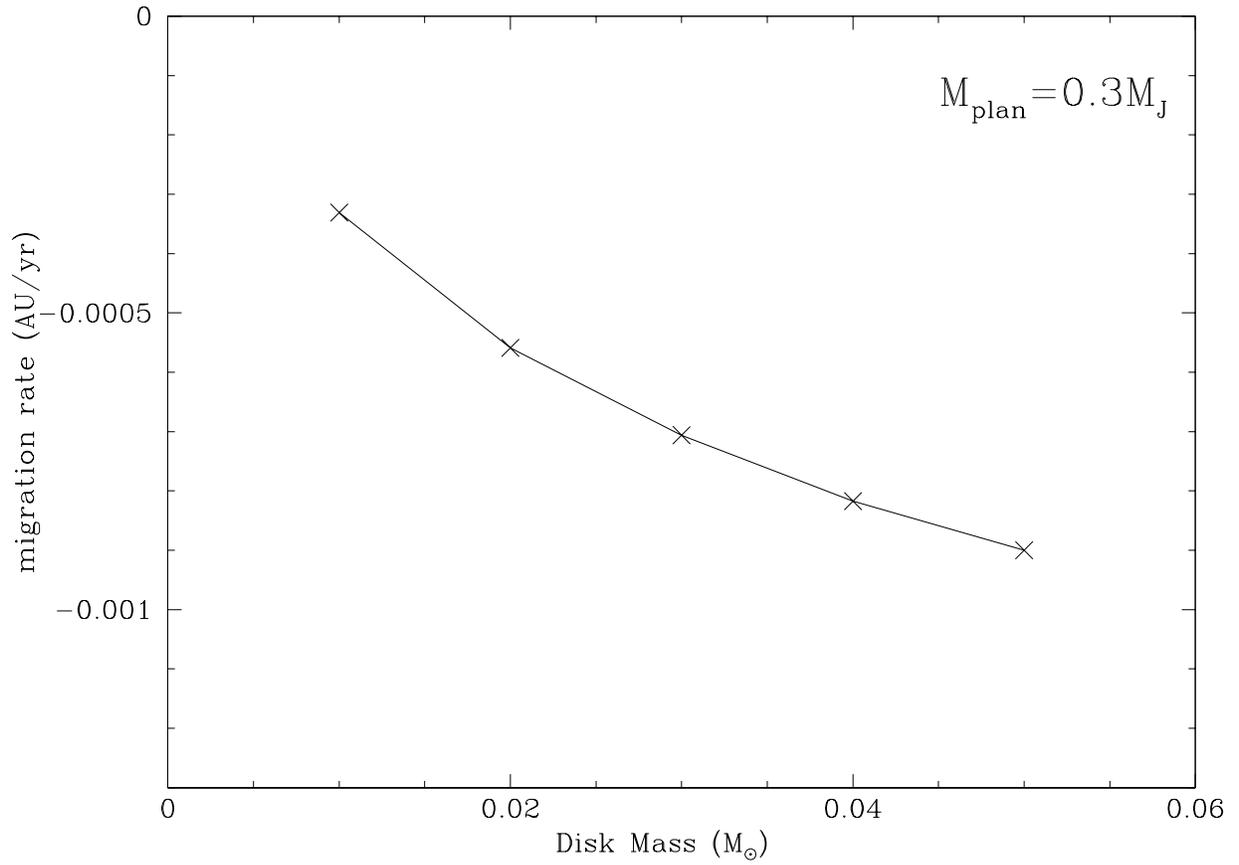,height=5in,angle=-90}
\caption[Disk mass dependence of migration rates]
{\label{fig:rates-disk}
The migration rate of a \mplan=0.3\mj\  planet through disks with
various masses. Symbols are as in figure \ref{fig:rates-mass}. For
each case, the planet was able to open a gap. }
\end{figure}

Such a massive disk as is assumed for the simulations in figure
\ref{fig:rates-mass} may only be appropriate early in the disk
lifetime, while the planet may grow to such large masses only much
later. Figure \ref{fig:rates-disk}, shows the migration rates
determined for a planet with mass 0.3\mj, but with lower disk masses,
between \mdisk=0.01 and 0.05\msun. We find a factor three speedup in
the migration rate occurs over a factor five increase in disk mass,
which while still appreciable is less than the linear dependence we
expect from equation \ref{eq:mig-timescale}. We note the
proportionality is shallower at higher disk masses than at lower,
perhaps indicating that the interactions have begun to saturate at
the higher disk masses. We conclude that the expected linearity is
only partially recovered in these models. 

In all cases, both for varying planet mass and for varying disk mass,
the fitted \tone\  migration rates are very rapid: more than 0.5~AU
per thousand years, and usually about 1~AU per thousand years. Planets
with masses greater than 0.3\mj\  begin with very rapid migration
rates, but within 3000~yr or less, build a deep gap where little disk
matter remains. This gap slows their migration rate because the large
dynamical torques acting on the planet decrease. If they remain in a
\tone\  migration phase, planets less than 0.3\mj\  would impact the
star in only a few $\times10^3$~yr. Lower disk masses will slow this
process and will reduce the planet mass sufficient to open a gap,
however an important point to note is that there is a critical region
in phase space in which the planet is vulnerable to rapid migration
and in which gap formation occurs too slowly to halt the migration. In
order to survive, a planet must gain mass quickly enough to traverse
this \tone\  migration region of parameter space more quickly than it
migrates through the disk.

For a 0.05\msun\  disk, the phase space of planet masses which undergo
continuous \tone\  migration region is defined from above by the
condition that the planet mass be less than 0.3\mj. The region will
also be defined from below in the sense that below some mass, a planet
will no longer couple strongly to the disk via gravitational torques
and its migration will be slow. The result shown in figure
\ref{fig:rates-mass} show that a 0.1\mj\  planet is already well above
this onset mass in a 0.05\msun\  disk. We have not attempted to
outline the parameter space further, either for planet masses below
0.1\mj\  or for lower disk masses, because of the limitations imposed
by our coarse resolution. At our standard resolution for example, the
Hill sphere of a 0.1\mj\  planet is about the same size as two grid
zones. As we will see in section \ref{sec:var-soft}, resolving the
mass distribution on this size scale is critical for determining the
true migration rate.

\subsection{The transition from \tone\  to \ttwo\  migration}
\label{sec:type1-2} 

The formation of a gap eventually causes the migration to slow and,
when it gets deep and wide enough, stop. This condition defines
`\ttwo' migration, where the migration is tied to the viscosity of the
disk rather than to dynamical processes like spiral wave generation
and gravitational torques (`\tone' migration). The arguments and
derivation of the \tone\  migration time scale in equation
\ref{eq:mig-timescale}, were also used in \citet{Ward97b} to determine
the conditions for the transition from \tone\  migration to \ttwo\
migration, at a mass he refers to as the `Shiva mass' whose definition
is:
\begin{equation}\label{eq:Shiva}
{{M_S}\over{M_*}} \approx 2 \alpha_{SS}^{2/3}
              \left({{M_*}\over{\Sigma r^2}}\right)^{1/3}
              \left( {{c_s}\over{r\Omega}}\right)^3,
\end{equation}
where $M_S$ is the Shiva mass, $\alpha_{SS}$ is the \citet{SS73}
viscous parameter and the radius, $r$, is position of the planet.
However, we saw in the last section that the predictions of equation
\ref{eq:mig-timescale} are only partially recovered in the simulations
and therefore cast doubts on the derivation of the Shiva mass as well.

Indeed, some discrepancy exists. Given the parameters of our
simulations (section \ref{sec:initconds}), we expect the Shiva mass to
be $M_S\approx 1.4\alpha^{2/3}$\mj, which requires a very large value
$\alpha\sim0.1$ to match the 0.3\mj\ transition mass seen in our
simulations. With more usual values of $\alpha=0.01$ or 0.001, we
derive $M_S\approx20$\me\  or 4.5\me, respectively, far lower than 
in our simulations. In order to explore the
discrepancy with theory, we determine empirically the approximate disk
conditions that define the boundary between \tone\  and \ttwo\  
migration for a given system as it evolves forward in time.

\begin{figure}
\begin{center}
\psfig{file=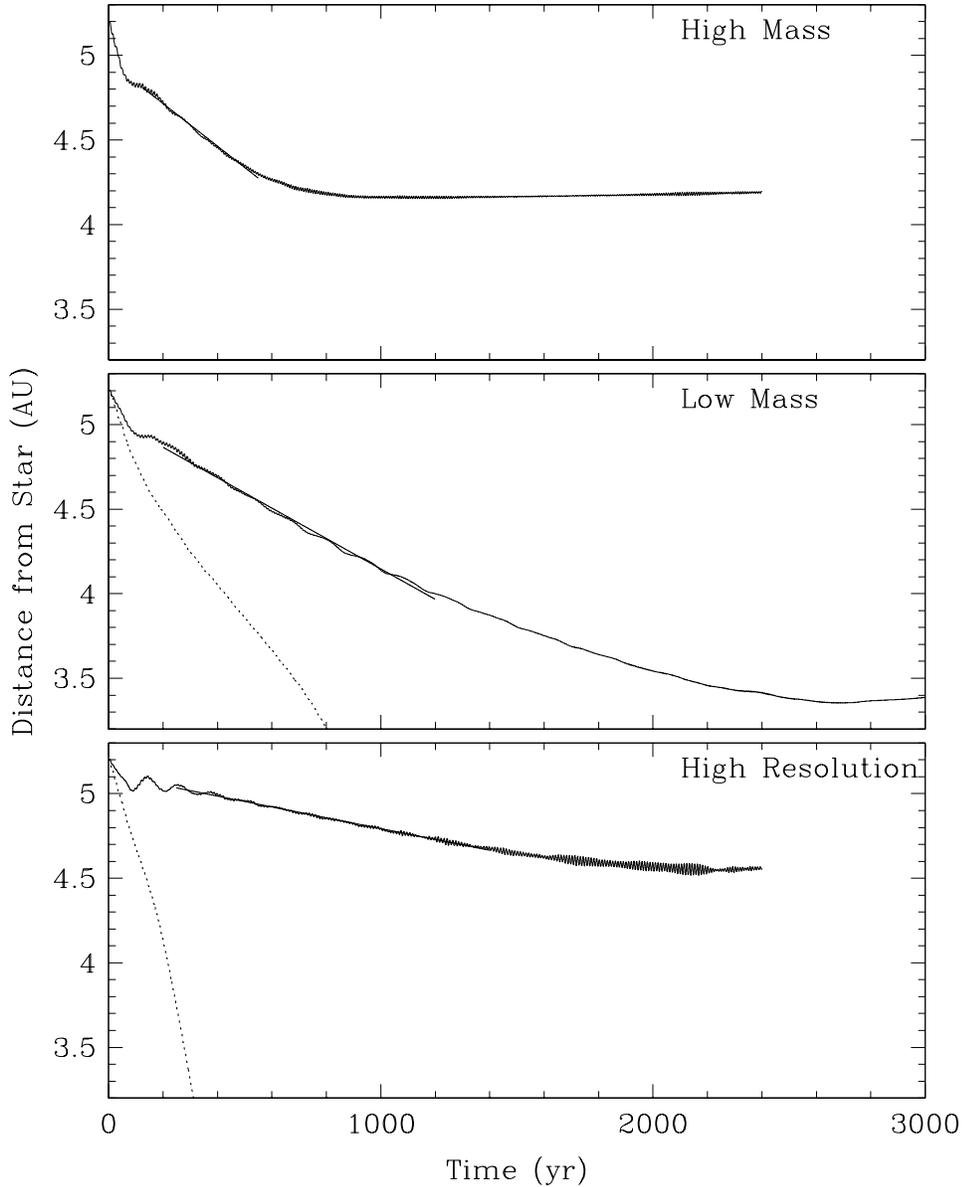,height=17.2cm,rheight=15.5cm} 
\end{center}
\caption[Distance of the planet from the star as a function of time]
{\label{fig:orb-migrate}
The evolution of the planet's orbit for the high mass, low mass
and high resolution prototype models (solid curves). For the
high mass prototype, rapid inward initial motion is caused by the
large torques exerted on the planet as it sweeps out matter and forms
a gap, but after a deep and wide enough gap develops little further
motion occurs. Neither of the lower mass simulations are able to form
gaps until much later in the simulations, and so continue their rapid
migration through the disk for much longer times. The dotted curves
refer to the simulations which omit disk self gravity. The solid
lines define linear fits to the trajectory during the period of time
during which the migration rate was near constant and the gap had yet
to form. Small and rapid inward/outward oscillations visible in the
motion are of much smaller magnitude than a single grid zone and are
probably not significant. Longer period ($\sim100$~yr) oscillations
visible in the low mass prototype are artifacts of the grid structure
on the trajectory.} 
\end{figure}

First, we attempt to gain some qualitative insight from figure
\ref{fig:orb-migrate}. Looking specifically for a moment at the high
mass prototype, we see that the migration begins to slow after about
5-600~yr of evolution, and essentially stops after $\sim900$~yr.
Comparison with figure \ref{fig:az-ave-dens-hi} shows that the planet
begins to slow it migration when the gap is 3~AU wide and has surface
density $\Sigma_{gap}\sim200$~g/cm$^2$. The transition is complete
and little further migration occurs when the system has evolved for
900~yr, and the surface density is $\Sigma_{gap}\sim 100$~g/cm$^2$.
We see similar behavior in the low mass prototype, except that the
time scale is extended to about 1800-2000~yr before the rates slows
significantly and 2700~yr for it to stop, with surface densities of
$\Sigma\sim3-400$~g/cm$^2$ and $\Sigma\sim300$~g/cm$^2$ at 2000 and
3000 years, respectively.

Disks with mass density at or below these values and planets as
massive as 0.3\mj\  will have already undergone a transition to
\ttwo\ migration, so that migration will be slow. The results of our
{\it `dis'} series of simulations did indeed confirm that 0.3\mj\
planets embedded in less massive disks (i.e. with
$\Sigma\sim2-400$~g/cm$^2$) migrate only a few tenths of an AU from
their initial orbital position before they form gaps. With a surface
density distribution like that in our simulations (i.e. $\propto
r^{-3/2}$), a value of 200~g/cm$^2$ at 4~AU corresponds to a disk
mass of \mdisk$\approx$0.01\msun\ between 0.5 and 20 AU. With a
shallower distribution ($\propto r^{-1/2}$) the implied disk mass is
\mdisk$\approx$0.02\msun, both of which are comparable to the minimum
mass solar nebula required to make the planets in our own solar
system. We conclude that minimum mass solar nebula disks with
embedded planets of 0.3\mj\  or larger will exhibit \ttwo\
migration. Lower mass planets may also successfully make the
transition to \ttwo\ migration in a minimum mass solar nebular as
well, however we have not studied that question in detail.

\begin{figure}
\psfig{file=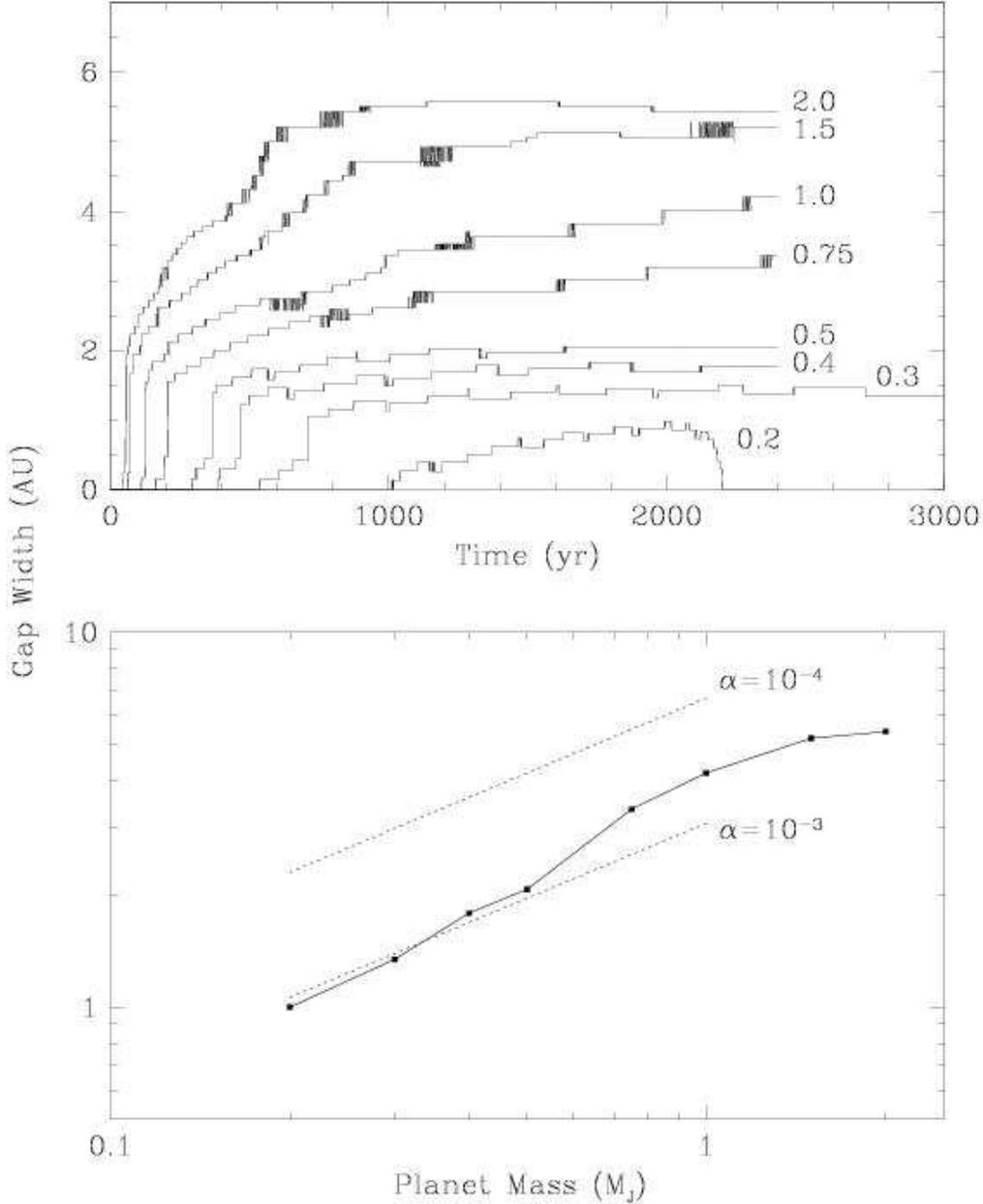,height=7.15in,rheight=6.75in}
\caption[Gap Width]
{\label{fig:gap-size}
The gap width for the {\it mas} series of simulations. In the top
panel the gap width is shown as a function of time as the simulation
proceeds. The curves are jagged because the gap width is measured as a
discrete number of grid zones. Each number shows the planet mass (in
\mj) corresponding to each curve. The bottom panel shows the final gap
width as a function of planet mass. The inner and outer gap edges are
defined as the radii at which the surface density drops to a factor of
two below the initial surface density. Except for the \mplan=0.2\mj\
model where we take the value at time t=2000~yr, the gap width is
defined as the width at the conclusion of the simulation, usually
2400~yr. The dotted curves define the gap width derived from eq.
\ref{eq:Tak-gap} and assuming a viscous coefficient
$\alpha_{SS}=10^{-3}$ and $10^{-4}$ as marked. } 
\end{figure}

A more quantitative measure of our results is the gap width as a
function of planet mass. In the top panel of figure
\ref{fig:gap-size}, we plot the width of the gap vs. time for each of the
{\it mas} series of simulations. Initially unperturbed disks will
form gaps within a few hundred years of the evolution, when perturbed
by the passage of a planet with mass $\ge0.3$\mj. After its initial
fully \tone\  phase, the gap forms and widens quickly to near its
final value, then (for the higher planet masses) slowly increases
further for the duration of the simulations. In addition, if we
compare figure \ref{fig:gap-size} to figure \ref{fig:orb-migrate}, we
see that the formation of a gap does not necessarily imply an
immediate halt to the migration. Simulations with planets $<0.3$\mj\
enter the transition to \ttwo\  migration temporarily, but do not
exit. Instead, they return to the fully \tone\  phase as they move
inward into higher density regions of the disk, finally impacting the
inner grid boundary. 

The analytical work of \citet[][hereafter TML]{Takeuchi96} predicts
that the gap width will be approximately
\begin{equation}\label{eq:Tak-gap}
\delta r \approx 0.9 a_{\rm pl} 
             \left({{q^2}\over{\alpha_{SS} h_0^2}}\right)^{1/3},
\end{equation}
where $q$ is the mass ratio between the planet and star and $h_0$ is
the dimensionless scale height of the disk, and assuming that the gap
is relatively narrow. Since we assume separate mechanisms for the
dissipation of waves and for viscous spreading (section
\ref{sec:diss-num}), rather than a single mechanism of constant
magnitude ($\alpha$ value) for all dissipation sources, correspondence
between our results and equation \ref{eq:Tak-gap} would appear to be
difficult to arrange. However, since we have assumed local damping, we
can neglect spatial variation in dissipation, allowing us to
substitute a single, unknown value for the dissipation cast as an
$\alpha_{SS}$ model. This will allow us to establish the veracity of the
theoretical model, given only correspondence with the proportionality.

In the bottom panel of figure \ref{fig:gap-size} we show the gap width
as a function of planet mass, measured at the end of each simulation
when the evolution and gap structure had reached their near steady
states. Also shown are the gap widths predicted by eq.
\ref{eq:Tak-gap}, assuming a scale height of $h_0=0.045$ and a planet
orbit radius of $a_{\rm pl}=$4.5~AU. Over the range between 0.2 and
1.0\mj, the gap width derived from the simulations is slightly steeper
than the predicted proportionality of \mplan$^{2/3}$. We believe that
the difference is due mainly to the spatial variation of our numerical
dissipation. Overall, the differences are small and the agreement
between theory and simulation is good.

Although we cannot specify a single input value of $\alpha_{SS}$, we
can use equation \ref{eq:Tak-gap} along with the results of our
simulations to provide a useful {\it a posteriori} check on the
overall background (non-wave) dissipation in PPM. For our standard
resolution, figure \ref{fig:gap-size} shows that the gap sizes
correspond to $\alpha\lesssim10^{-3}$. While this value puts our
simulations well within the range of values in general use for
circumstellar disks, it is in conflict (by a factor of 100) with the
value estimated from equation \ref{eq:Shiva} and the 0.3\mj\  critical
gap opening mass obtained from the same simulations.

The origin of the conflict lies in our separation of wave dissipation
from other forms. The gap width $\alpha_{SS}$ value (of $10^{-3}$) is
based on the long term state of the system after waves have mostly
decayed away. It therefore reflects the low ambient internal
dissipation expected of accretion disks (section \ref{sec:diss-num})
and provided by PPM. On the other hand, the Shiva mass from equation
\ref{eq:Shiva} is derived from a much more active part of the
evolution, where waves generated by the planet reach significant
amplitudes for which shocks may form. As we noted in section
\ref{sec:diss-num}, shocks are known to develop near the planet for
planet's more massive than $\sim30$\me, so the dissipation there will
be high for quite physical reasons, and correspond to a locally high
value of $\alpha_{SS}$, for the times when those shocks are present.

While the strong wave dissipation implied by our local damping
assumption may indeed be responsible for the two different values of
the Shiva mass, that responsibility cannot be extended to the
migration rate's lack of sensitivity to changes in the planet mass. In
that case, local damping neither helps nor hinders the migration until
large changes in the density distribution have evolved. Since we limit
our fits to the \tone\  evolution period, they will not be strongly
affected by the details of the wave damping and the conflict with
theory remains.  

\subsection{The strength of dynamical interactions between the planet,
its envelope and circumstellar disk}\label{sec:var-soft}

While we find very good agreement for the gap width to planet mass
proportionality, the same cannot be said of the migration rate
proportionalities, which were reproduced in one case and not another.
Although we could show at least a close to linear dependence of the
migration rate on disk mass, we did not find the expected linear
dependence on planet mass. Instead the rate varied by less than a
factor two over a factor 20 change in planet mass. Both the
predictions of the migration rates and the gap widths are derived from
the same theoretical framework, namely that the gravitational torques
are generated from Lindblad resonances. Why does the correspondence
fail in some cases and not in others? 

When a gap has already formed and little matter remains near the
planet, the assumption underlying equations \ref{eq:mig-timescale} and
\ref{eq:Tak-gap} (that only Lindblad resonances are important) may
indeed be valid, especially if the gap is wide and deep enough that
interactions close to the planet are small. In the immediate vicinity
of the planet, especially during the \tone\ phase, other interactions
like corotation resonances and mass accretion may be important, each
giving much larger significance to the small scale mass distribution
than is assumed using equations \ref{eq:mig-timescale} and
\ref{eq:Tak-gap}.

How influential are the mass structures that form around a planet in
terms of their effect on the dynamical evolution of the planet/disk
system as a whole? We study this question by varying the gravitational
softening parameter (see eq. \ref{eq:pm-force}) used to calculate the
force between the planet and disk. With large softening, little
significance is given to the matter near the planet, while with small
softening much more significance is given. We vary gravitational
softening radius between 0.1 and 4.0 times the size of one grid zone,
corresponding to a physical size between 0.015 and 0.6~AU at our
standard resolution of 128$\times$224 grid cells. For these
simulations (designated {\it sof} in table \ref{tab:table-mig}), we
assume the same initial conditions as for our low mass prototype
simulation above.

\begin{figure}
\psfig{file=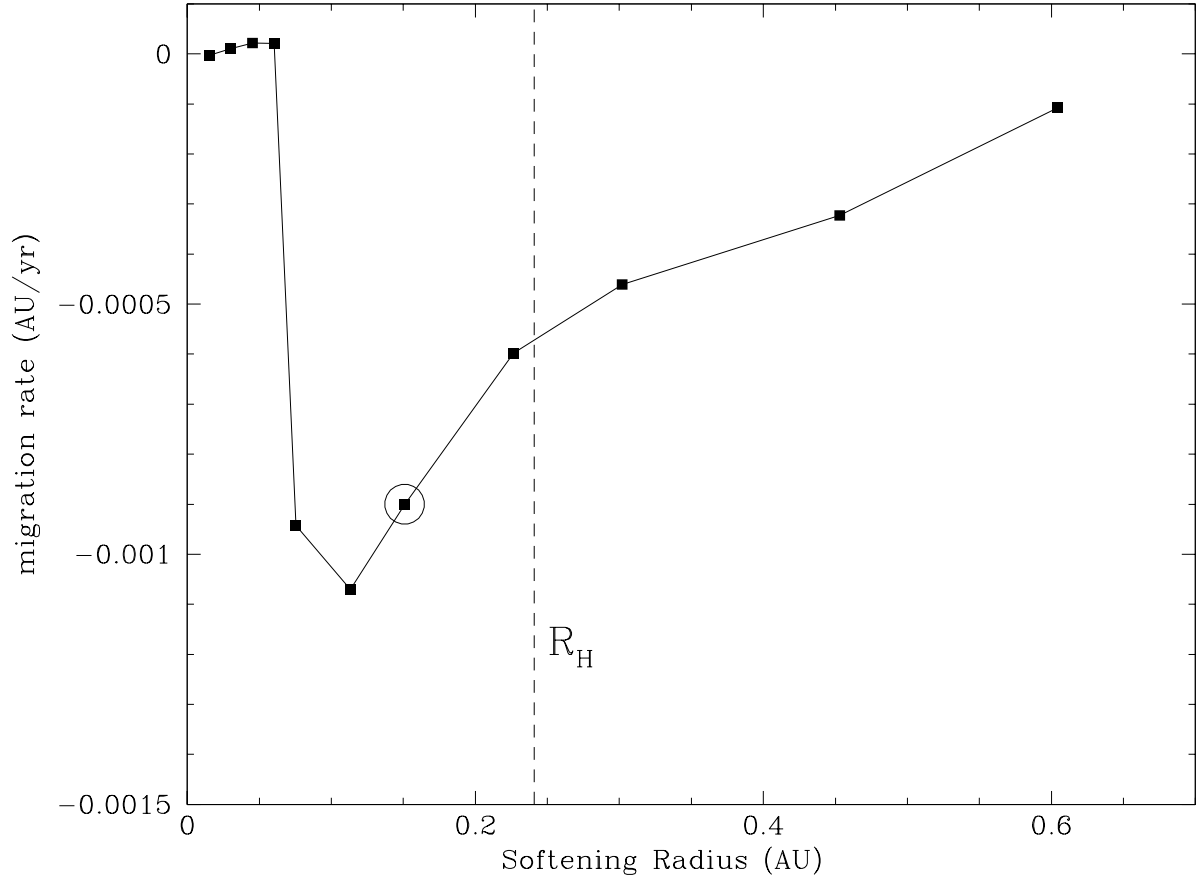,angle=-90,height=5in}
\caption[Softening parameter dependence of migration rates]
{\label{fig:rates-soft}
Variation of the migration rate with different numerical softening
parameters for a 0.3\mj\  planet. The circled point is the simulation
{\it mig3}, for which the softening parameter is $\epsilon=1.0$. The
vertical dotted line is the Hill radius of the planet at 5 AU, and is
also the scale height of the disk there. For runs with $\epsilon<0.5$,
the migration is near zero, which is a consequence of the numerical
suppression of the migration by the tendency of the planet to become
locked to a single azimuthal ring of grid cells. } 
\end{figure}

The migration rates obtained from these simulations are shown in
figure \ref{fig:rates-soft}. The migration rate increases by a factor
of about five as the softening decreases from 0.5~AU to 0.1~AU. The
largest increases occur as the softening decreases to the size of the
Hill radius or smaller. As the softening decreases below
$\epsilon=0.5$, the magnitude of the migration rates decreases to
zero, as the planet becomes unphysically bound to a single ring of
grid zones.

For the physically relevant range of softening parameters
($\gtrsim0.1$AU) studied, we can make the important conclusion that
the distribution of disk matter within 1--2\rh\ (i.e. with both very
similar orbit radius and orbital phase as the planet) of the planet
plays a critical role in determining its orbital evolution and fate
during its \tone\  migration phase. This is a stronger conclusion than
that made by one dimensional analyses, which are able to conclude only
that the radial region within a few \rh\  is important. We base it
both on the large increase in migration rates with decreasing
softening and our earlier (section \ref{sec:migration}) finding that
the migration rate is sensitive to the grid resolution employed in the
simulations. Although our simulations are two dimensional, the
conclusion is general in the sense that it includes both the two and
three dimensional mass distribution, since the softening affects
torques on a spatial scale very similar to the disk scale height.

\section{Discussion and Comparisons to Other Work}\label{sec:summary}

In this section we attempt to compare our results with some of what
has gone before, to place them in context. In terms of the physical
model, while we have employed a deliberately simplistic thermodynamic
treatment, we know of only one study other than our own \citep{MVS87},
after GT79 which includes disk self gravity in an analysis of planet
migration. Even there, the discussion is of the properties of wave
generation and propagation of disturbances in the disk, rather than on
the consequences for the planet. Explicitly stated or not, the
assumption made in studies of planet migration so far has been that
self gravity would contribute only negligibly to the migration since a
disk with as low a mass as ours should be stable (i.e. Toomre $Q>>1$).
Given its dramatic effect in our simulations, this assumption must be
discarded in future analyses.

What is the origin of the large rate changes when disk self gravity is
included? Although the answer to this question is beyond the scope of
our work here, where we have made only relatively qualitative
comparisons of our results to analytic predictions, we may be guided
in our speculations by \citet{Ward97a}. He shows that pressure forces
will cause slight shifts in the rotation curve of the disk, relative
to purely Keplerian flow, causing the resonance locations themselves
to shift. These shifts can be responsible for very large changes in
the net Lindblad torque on the planet because that torque is a sum of
two large terms of opposite sign. Disk self gravity will produce
similar shifts and we will find in \jtwo\ that in fact resonance
shifts due to inclusion of disk self gravity can cause large changes
in the gravitational torques, so they may be responsible for at least
some of the observed sensitivity. 

Previous numerical studies of disk with embedded planets fall into two
general categories. First, there are studies like our own that begin
with an unperturbed disk containing a planet
\citep[e.g.][]{Bryden99,NPMK} and second, studies that begin with an
already formed gap around the planet \citep{LSA99,Kley99,KDH}. In
general, the previous studies have focused on planets with mass 1\mj\
or larger and a single disk mass (a `minimum mass solar nebula'),
during the gap clearing process and following \ttwo\ migration epoch.
In many instances they suppress planet migration in order to explore
gap formation and mass accretion processes in isolation. The period of
gap clearing is common to both our study and the previous studies and
it is on this basis that we compare our results to theirs. 

\subsection{Comparisons to semi-analytic and numerical `$\alpha_{SS}$'
models}\label{sec:alpha-comp}

We find that planets more massive than $\sim 0.3$\mj\  can open a gap
sufficiently wide and deep to halt their dynamical migration through
the disk (i.e. transition to \ttwo\  migration), even for relatively
massive (\mrat=0.05) disks. The critical planet mass of 0.3\mj\  agrees
very well with a result from \citet{PP3_LP} that only planets with
\mplan/\msun$>3(H/r)^3$ could form gaps, but provides the additional
information that there is enough time for the gap to open before the
planet is lost by accretion onto the star, even if the system starts
from the condition that no gap has yet started to form.

The gap sizes produced in our simulations are somewhat wider than the
values quoted by \citet{Bryden99} (a half width of $\sim0.2a_{\rm pl}$
for a 1\mj\  planet, compared to our value of $\sim0.5a_{\rm pl}$),
mainly because of a difference in the definition of a gap. While both
the TML and \citet{Bryden99} derivations follow from the same physical
arguments and produce the same proportionalities, it is not clear what
factor should be used in order to produce an equality, because of the
ambiguity about what point in the evolution that equality should
signify. The usefulness of a gap width relation such as equation
\ref{eq:Tak-gap} is therefore limited. While we used the TML
definition in section \ref{sec:type1-2}, it is clear that forming a
gap of a given width and the end of rapid migration are not equivalent
statements. On the other hand, the \citet{Bryden99} definition
produces a time scale for gap opening proportional to the width to the
fifth power, so a small error in the width (through e.g. a
mis-estimation of $\alpha$ or the disk scale height) will have large
consequences on the predicted time scale, perhaps leading to the
incorrect conclusion that a planet might or might not be accreted by
the star when in fact the opposite result is the case. 

On one level, our result that the critical planet mass for opening a
gap is $\sim0.3$\mj\  is interesting because it validates the initial
conditions of the many models beginning with a massive planet and an
already formed gap. However, it does not address the question of how
that initial condition came to pass--lower mass planets do not form
gaps and instead continue to migrate rapidly. Unless some mechanism
exists by which they can avoid such rapid migration, they will be
accreted by the star, completing the `Shiva' scenario of
\citet{PP4_WH}. Moreover, the Shiva mass is derived while neglecting
disk self gravity, while our simulations include it. While we have not
investigated the exact transition mass in simulations of non self
gravitating disks, it is clearly much higher than 0.3\mj, making its
value even more inconsistent with the theory. 

The gap opening mass from our simulations is inconsistent with not
only the predicted Shiva mass of $\lesssim10-20$\me\  discussed in
\citet{PP4_WH} but also the value obtained from estimating an
$\alpha_{SS}$ value from the gap widths themselves (figure
\ref{fig:gap-size}). This conflict exposes an important shortcoming in
many analytic theories of migration. In nearly all such models, the
magnitude of dissipation (ordinarily a single $\alpha_{SS}$ value)
defines all of the dissipation in the system at every point,
effectively separating the assumption of local wave damping from the
magnitude of the dissipation supposed to cause it. Given the
importance of spatially and temporally varying dissipation processes
like shocks in models of planet migration, this separation is quite
dubious. In our models, we have retained both a low ambient
dissipation and the local damping assumption in a self consistent
manner, leading to the conflict. In order to retain self consistency,
analytic theories must also allow for this separation.

\subsection{Sensitivity of the results to spatial
resolution}\label{sec:reso-concl}

Based on the partial agreements and disagreements with theory (section
\ref{sec:var-soft}) and on the high sensitivity to the mass
distribution (i.e. the resolution and the gravitational softening), we
made the physical conclusion that the two and three dimensional mass
distribution within 1--2\rh\  of the planet is very important for
determining its fate.

In terms of the radial distribution of matter, requiring a very well
resolved mass distribution is far from a new conclusion. A large
fraction of the torque comes from higher order ($m\sim10-20$) Lindblad
resonances located about a disk scale height inward and outward of the
planet. In part, the added significance in this work is that the
sensitivity extends to the azimuth and vertical coordinates as well. A
previous shearing sheet calculation \citep{Miyoshi99} showed that
allowing for three dimensional structure near the planet can cause a
factor $\sim2.5$ decrease in the magnitude of the torques relative to
two dimensional calculations, but because their calculations modeled
only a small region around the planet it was unclear how the results
might translate into more global models. A contrasting result with
global 3d simulations \citep{KDH} found little difference between
rates determined from 3d calculations and from previous two
dimensional versions.

In our simulations, the 1--2\rh\  spatial scale is resolved by only a
few zones, which gives a relatively coarse picture of the mass
distribution there. Indeed, the sensitivity of our results to
resolution is a major factor in making the conclusion in the first
place. Given the variation in migration rates for physically identical
simulations of different resolution, we must conclude that our
resolution is not high enough to determine a well converged value for
the migration rates of the planets. The resolution employed in our
models is quite similar to that employed in many other works
(typically between 1 and 25 $\times10^{-3}$~AU$^2$, though with grid
spacings that vary substantially between works), and we would expect
that many of them are affected by the same factors affecting our own.

What resolution is sufficient? The models of \citet{LSA99} resolve the
Hill sphere of a 1\mj\ planet in two dimensions with about 250 zones,
using a variably spaced grid so that resolution close to the planet is
high. This is a factor $\sim 5-6$ better spatial resolution of area than
our high resolution models (which resolve the Hill radius of a 0.3\mj\  
planet with $\sim20$ zones), corresponding to a factor $\sim2.5$ better
linear resolution. They find little variation in the flow pattern around
the planet with varying softening parameter, so it seems likely that
their resolution is sufficient to resolve the gravitational torques as
well. Therefore, unless the migration rates change by a large factor
with an additional factor $\sim2-3$ in linear resolution beyond our high
resolution models, our rates will also not suffer from a large error.

Since many of the concerns regarding an incompletely converged
migration rate are relevant both to this paper and to \jtwo, we will
address the veracity of conclusions made in both papers here. In most
cases, although undesirable, the lack of full convergence does not
greatly diminish the value of our conclusions because they have been
based not on the value the migration rate, but rather on its variation
with various physical and numerical parameters. While a systematic
error in the magnitude of various quantities (gap width and migration
rate as a function of planet mass, or migration rate as a function of
disk mass) may be present, the conclusions regarding the
proportionalities remain unaffected because each simulation in each
series was done at the same resolution, meaning that any systematic
effects will affect each in the same way. Our finding that disk self
gravity was important for the evolution will also be unaffected, as
will our finding in \jtwo that that the rate of accretion onto a
planet is dominated by the small scale conditions very close to it
rather than by the large scale disk morphology. 

The remaining concerns are with respect to the magnitudes of the
migration rates (or equivalently, the torque magnitudes) and through
them, of quantities like the Shiva mass and the mass accretion rates
required for the planet's survival. Two conclusions may be affected by
such variation. First, the conflict between torques from theory and
simulation (\jtwo) become smaller if the migration is in fact faster.
Since the migration rates and torques instead decreased with higher
resolution (at least for the situation most relevant to real
systems--with disk self gravity included), the torque magnitude
conflict either remains or is understated in our work.

Second, with a slower migration rate, the Shiva mass and the lower
bound of the accretion rate onto the planet derived from that rate in
\jtwo would each be lower. Based on these two values, we made the
conclusion that the early stage of planet formation was inconsistent
with both a core$+$circumplanetary disk and a static spherical
envelope morphology. If the true \tone\  migration rates are a factor
ten slower than those shown in figure \ref{fig:rates-mass} (i.e.
$\sim10^{-4}$~AU/yr), the conclusion will remain valid, though
weakened, since the statement was based upon a very conservative model
for the planet's disk (that its mass was relatively small). It would
be nullified in the unlikely case that the true \tone\  rates were a
factor of one hundred slower ($\sim10^{-5}$~AU/yr).

\acknowledgements
We would like to thank Willy Kley and Pawel Ciecielag for many
productive conversations during the evolution of this work, Bill Ward
for a very helpful discussion about saturation of the torques and Jan
Alibert for a careful reading of the manuscript. WB acknowledges
partial support from the Swiss National Science Foundation. AFN is
grateful for financial support from the UK Astrophysical Fluids
Facility (UKAFF), during the final months of the preparation of this
manuscript.

\appendix

\section{Numerical issues concerning the point mass force
calculation}\label{app:pm-force}

In this appendix, we examine the numerical influences of the softening
radius and grid resolution on the force calculation responsible for
determining the planet's trajectory, in order to understand any
systematic errors that we may make and the requirements needed to
obtain reliable results. The gravitational force between the planet
and the disk must be modified from its true form because the matter in
the disk is resolved only as a set of small zones in a grid. Without
modification, a close encounter with the center of a single zone will
cause an effectively infinite and unphysical force to be calculated.
Further, when the trajectory of the point mass is along a grid
direction, the force calculation will be systematically slightly
different when it is located near the zone centers, compared to that
when it is near the zone interfaces. 

In order to avoid these kinds of errors, we would like the region over
which the force is modified to be large, so that any single grid zone
(or azimuthal ring of zones) does not produce an unduly large
influence on the trajectory. On the other hand, the physical effect of
a large softening parameter is to blur or `turn off' the mutual
interaction between the planet and disk exactly where it may be the
strongest, and may lead to an incorrect model of the evolution.
Therefore, we would also like the modified region to be as small as
possible in order to correctly calculate the force due to the disk
mass close to the planet.                                                   

To address both of these requirements, we use direct summation of the
force between the planet and each grid zone using the Plummer force
law as defined in eq. \ref{eq:pm-force}. This form allows the
softening radius, $\epsilon$, to be chosen as a constant fraction of
the zone size that the planet is in, but still allows the freedom to
choose the value of that fraction differently in different
simulations. We also experimented with a constant softening parameter,
but found no difference between those results and with those using a
constant fraction of the current zone size, since the force is
proportional to $1/\epsilon^2$ and radially adjacent zones change in
size by $<$2\%. 

\begin{figure}
\begin{center}
\psfig{file=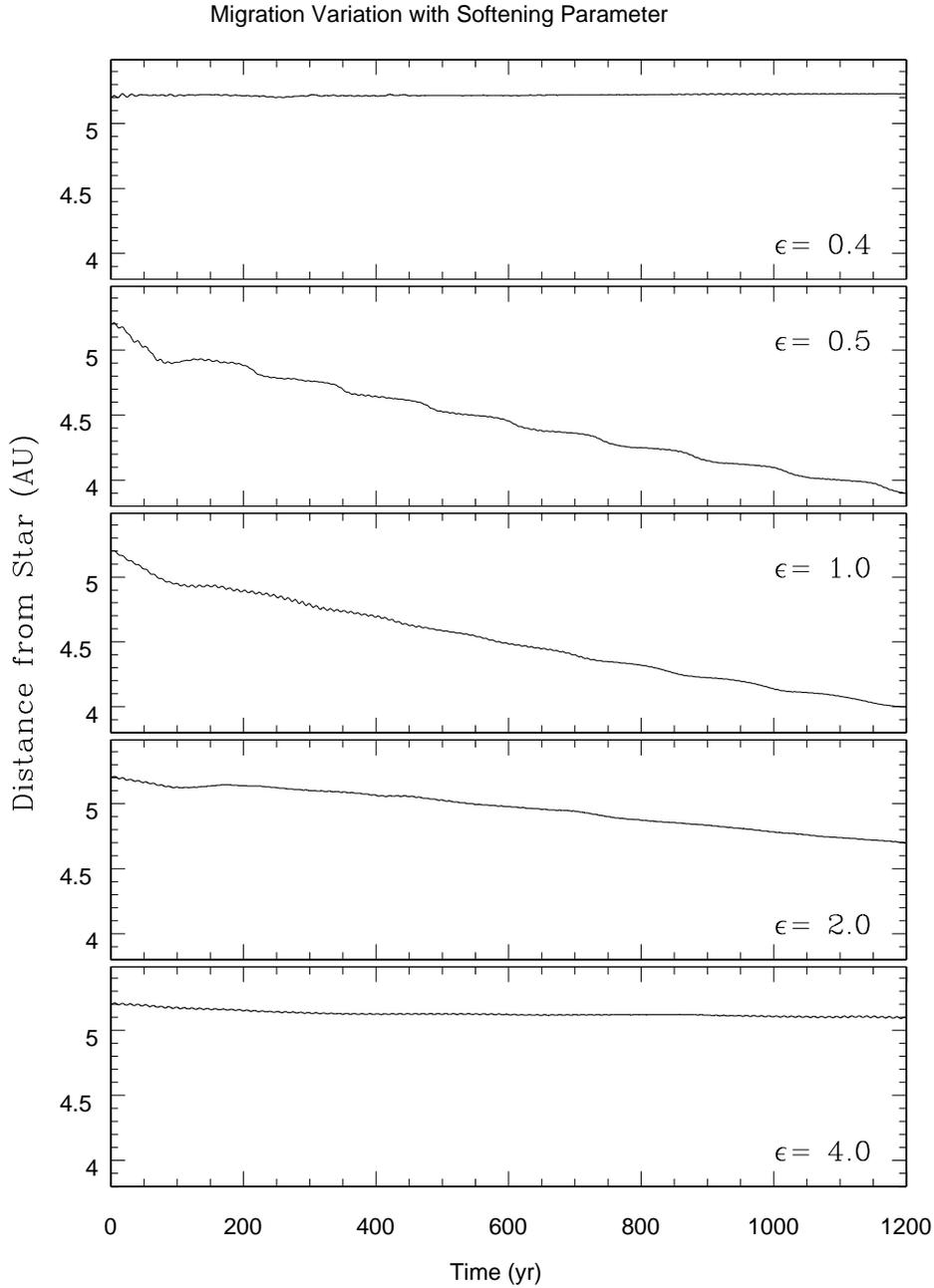,height=7.0in,rheight=6.5in}
\end{center}
\caption[Variation with Softening]
{\label{fig:var-migrate}
The evolution of the orbit of the planet with different softening
parameters, $\epsilon$ (given in units of the local grid zone size),
for the gravity due to the planet. For very large softening (e.g.
$\epsilon\ge2$), the simulations produce very slow migration rates.
Migration proceeds more rapidly for intermediate softening values,
($1.0\lesssim\epsilon\lesssim2.0$). but for still smaller values
($\epsilon=0.5,0.75$), the grid structure produces a `stair step'
pattern in the trajectory which becomes more noticeable for smaller
softening. For the smallest softening, the planet becomes locked to a
single azimuthal ring of grid cells and the results become
unphysical.}
\end{figure}

We show the trajectories of the planets in a series of simulations
that varied the softening parameter in figure \ref{fig:var-migrate}.
With very large softening values we find that the planet migrates only
very slowly through the disk. In this case, only relatively distant
parts of the disk are able to influence the trajectory of the planet
since the softening strongly suppresses the gravitational forces from
nearby material. As the softening value decreases, the migration rate
increases as the forces due to the disk mass close to the planet
approach their unsoftened values. However, this influence ultimately
becomes incorrectly modeled because of the finite resolution of the
grid. For softening values $\epsilon\sim0.5-0.75$, the radial
trajectory shows clear evidence of a stair step pattern, as the planet
passes inward into the influence of successive rings of grid zones.
When the softening value reaches less than half the size of one grid
zone, the influence of a single grid zone or a single azimuthal ring
of grid zones dominates the gravitational force on the planet,
unphysically binding its orbit to that ring so that its migration rate
drops to zero. 

For simulations with higher mass planets or higher grid resolution,
the situation reverses: The migration rates obtained when this occurs
are very rapid, generally faster than 1~AU/100~years. The most massive
planet we simulated ({\it ms10}, 4\mj) migrated inward to half it's
initial orbital radius in 60~years, then stopped due to the formation
of a gap and the build up of disk matter between it and the inner grid
boundary. This behavior is not the inertial migration regime as
discussed by \citet{HourWard84} and \citet{WardHour89}, but rather a
numerical effect brought about by the unphysical concentration of
large amounts of matter in only one or a few grid zones very near the
high mass planet, as it tries to form an `atmosphere'. This
unphysically high density causes the force on the planet to be much
greater than it would otherwise be, and is caused by our isothermal
equation of state assumption which implies efficient cooling. Such
assumptions break down when the optical depth becomes very high and
shock and/or compressional heating become significant. We would expect
each of these phenomena to occur near the planet.

While such effects are not unexpected, it is important to determine
the specific pathologies and the softening values that trigger them
(so as to avoid them) in our models and in our code. From these
results we conclude that a Plummer softening parameter smaller than
half the linear dimension of one grid zone should never be used, and
that values as small of 0.75 times the size of one grid zone can be
used without serious numerical defects in the simulation results for
low mass planets, but at least a value of at least 1.0--1.5 should be
used for higher mass planets or higher resolution simulations. In this
work we will typically use a softening radius of 1.0 times the size of
one grid zone.

\section{The numerical dissipation in PPM and its application to 
disk simulations}\label{app:ppm-diss}

Dissipation is present in all numerical schemes for solving the
hydrodynamic equations because small scale inhomogeneities are only
coarsely and inaccurately resolved. In calculations done on a grid for
example, unphysical mixing may occur due to the incorrect advection of
mass from one zone to another. Further, no wave shorter than twice the
length of one grid zone (i.e. the Nyquist wavelength) can exist, and
physical effects which depend on such waves will not be included in
the model. Longer waves may be present but will experience dissipation
as the system evolves forward in time because of errors in the 
assumptions in the numerical method used to model the system.

\citet{PortWood94} have quantified the numerical dissipation for
disturbances propagating through a grid as modeled by a PPM code.
They find in an empirical study that it is steeply dependent on the
ratio of the wavelength to the grid spacing. They have quantified the
decay rate of a sinusoidal shear flow traveling diagonally through a
two dimensional mesh and produce an empirical result that the decay
rate is proportional to the third and fourth powers of the shear
wavelength, $\lambda$, over the grid cell size, $\delta x$:   
\begin{equation}\label{eq:PW94-eq}                            
{{\dot a}\over{a}} = - \left[                                 
         A_s\left({{\lambda}\over{\delta x}}\right)^{-3} +    
         B_s\left({{\lambda}\over{\delta x}}\right)^{-4}\right]
             {{u_0}\over{\lambda}}                            
\end{equation}                                                
where $u_0$ is the initial velocity amplitude of the shearing wave
through the grid. $A_s$ and $B_s$ are the empirically derived 
coefficients reproduced in Table \ref{tab:table-PW} from      
\citet{PortWood94}. For different advective Courant numbers (defined
as $C=\delta t/\delta x \times v$, where $\delta t$, $\delta x$ and
$v$ are respectively the time step, the grid size and the fluid
velocity through the grid). Dissipation for a fluid shearing in a
direction parallel to the grid direction will be lower.       

In the context of Jovian planet migration, we expect from analytic
studies that waves of many different wavelengths will be excited by
the passage of the planet through the disk, and propagate in both the
radial and azimuthal directions. The azimuthal wavelength of a given
wave will be $\lambda_{az}=2\pi r/m$, where $m$ is the azimuthal
wavenumber. The wavelength of the radial component will be a function
of radius and will be dependent on the local conditions, relative to
the resonances in the disk. In general, the radial component of a
given wave will have much shorter wavelength than the azimuthal
component and be less well resolved. (Note that this is in essence a
statement of the `tight winding approximation'). Therefore the radial
component of the wave will contribute the dominant source of the
numerical dissipation of the combined pattern. Following the
discussion in \citet{PP3_LP}, the damping of radial waves in the disk
will be proportional to $e^{-\nu k^2t}$. Stated another way, the
logarithmic damping rate of the wave amplitude can be expressed as 
\begin{equation}\label{eq:viscampdot}                         
{{\dot a}\over {a}} = -\nu k^2,                               
\end{equation}                                                
where $k$ is the radial wave number, $k=2\pi/\lambda_r$.      

We can combine equations \ref{eq:PW94-eq}, \ref{eq:viscampdot} and the
standard \citet{SS73} expression, $\nu=\alpha_{SS}c_sH$, into a single
expression for the expression for the $\alpha_{SS}$ parameter which
quantifies the numerical dissipation of wave structures in disk
simulations inherent to the PPM algorithm. If we also assume that
$u_0\approx V_\theta$, we obtain $\alpha_{PPM}$ as:             
\begin{equation}\label{eq:ppm-alpha}                          
\alpha_{PPM} = {{\lambda r }\over{(2\pi)^2 H^2}}\left[        
              A_s\left({{\lambda}\over{\delta x}}\right)^{-3} +
              B_s\left({{\lambda}\over{\delta x}}\right)^{-4}\right].
\end{equation}                                                
Our standard resolution simulations have $r\delta\theta = \delta
r\sim0.12-0.15$~AU near the planet so that a wave with
$\lambda_{radial} \sim 1$~AU will be resolved with $\sim6-8$ zones.
Assuming conditions appropriate for our simulations, we derive a value
of $\alpha_{PPM}\sim 10^{-1}$, for a wave with $\lambda=1~AU$ located
5~AU from the star. At double this resolution (or equivalently, for
waves twice as long), we derive a value $\alpha_{PPM}\sim 10^{-2}$.
For the spiral waves excited by the planet, we therefore expect that
wave components with $<$1~AU wavelengths (roughly equivalent to
patterns with $m>10$) will be completely dissipated very near their
origination. 

Given equation \ref{eq:ppm-alpha}, we see that the dissipation is very
steeply wavelength and cell size dependent. Changing either by a
factor two changes the numerical wave dissipation by a factor of
$\sim10$. In common usage however, the $\alpha$ parameter takes one or
occasionally two values over the entire disk, making clear that the
dissipation in PPM and $\alpha$ models, as normally implemented, have
little in common. Further, equation \ref{eq:PW94-eq} (and therefore
also equation \ref{eq:ppm-alpha}) is strictly valid only for isolated
single wave forms. Dissipation of multiple superposed waves cannot be
calculated simply as a sum of the dissipation of individual components
and instead may experience little or no numerical dissipation,
depending on the local flow. A conservative approach therefore
requires that we use equation \ref{eq:ppm-alpha} as a guide rather
than a specification.

Infinite wavelength features (i.e. axisymmetric structures) will be
only weakly damped, but nevertheless remain subject to numerical
mixing and diffusion. We expect numerical diffusion is small in PPM,
because of the non-linear switches to detect impending diffusive
behavior in the flow and correct it. Indeed, \cite{FMA91} showed that
sharp features (e.g. shocks or contact discontinuities) will spread to
a width of only about two zones in a one dimensional flow before PPM's
discontinuity detection switches become active and prevent further
spreading.         

A gap in the surface density distribution is defined by two such
features, one interior to the planet and one exterior to it. The
requirement that the grid be large (radially) to provide sufficient
distance for spiral waves to dissipate before encountering a boundary
affords us relatively low resolution in the gap region. The gap
structure may therefore be affected by the numerical diffusion. We
found in section \ref{sec:type1-2}, that the gap width for a 0.3\mj\
planet is about $1.5$~AU, so that it is resolved in our standard
resolution runs by $\sim10$ zones. This means that nearly half of the
gap may be influenced by numerical diffusion. The same gap width in
our high resolution models would be resolved with twice as many zones,
while the numerical diffusion effects will remain limited to regions 2
zones wide. Therefore numerical diffusion and the tendency for gap
closing it implies will be lower in these runs. Looking specifically
at the results of section \ref{sec:type1-2}, we note that even when
resolved with only $\sim10$ zones (for our lower mass planets), the
gap widths obtained from the simulations imply a numerical diffusion
whose magnitude corresponds to $\alpha\lesssim10^{-3}$, assuming the
validity of the calculation for the gap width.

\clearpage

\singlespace
\begin{deluxetable}{lccccc}
\tablewidth{0pt}
\tablecaption{\label{tab:table-mig} Initial Parameters For Simulations}
\tablehead{
\colhead{Name}  & \colhead{Resolution} & \colhead{Disk Mass} &
\colhead{Planet Mass}  & \colhead{Softening} & \colhead{Duration} 
\\
\colhead{}  & \colhead{($r\times\theta$)} & \colhead{$M_{\odot}$} &
\colhead{$M_{J}$}  &  \colhead{$\epsilon$} & \colhead{(yr)} }

\startdata
mas1 & 128$\times$224 & 0.05 & 0.10     &  1.0\phn &    2400 \\
mas2 & 128$\times$224 & 0.05 & 0.20     &  1.0\phn &    2400 \\
mas3\tablenotemark{1} 
     & 128$\times$224 & 0.05 & 0.30     &  1.0\phn &    3000 \\
mas4 & 128$\times$224 & 0.05 & 0.40     &  1.0\phn &    2400 \\
mas5 & 128$\times$224 & 0.05 & 0.50     &  1.0\phn &    2400 \\
mas6 & 128$\times$224 & 0.05 & 0.75     &  1.0\phn &    2400 \\
mas7 & 128$\times$224 & 0.05 & 1.0\phn  &  1.0\phn &    2400 \\
mas8 & 128$\times$224 & 0.05 & 1.5\phn  &  1.0\phn &    2400 \\
mas9 & 128$\times$224 & 0.05 & 2.0\phn  &  1.0\phn &    2400 \\
ms10 & 128$\times$224 & 0.05 & 4.0\phn  &  1.0\phn & \phn600 \\
dis1 & 128$\times$224 & 0.01 & 0.30     &  1.0\phn &    2400 \\
dis2 & 128$\times$224 & 0.02 & 0.30     &  1.0\phn &    2400 \\
dis3 & 128$\times$224 & 0.03 & 0.30     &  1.0\phn &    2400 \\
dis4 & 128$\times$224 & 0.04 & 0.30     &  1.0\phn &    2400 \\
sof1 & 128$\times$224 & 0.05 & 0.30     &  0.1\phn &    1800 \\
sof2 & 128$\times$224 & 0.05 & 0.30     &  0.3\phn &    1800 \\
sof3 & 128$\times$224 & 0.05 & 0.30     &  0.5\phn &    1800 \\
sof4 & 128$\times$224 & 0.05 & 0.30     &  0.75    &    1800 \\
sof5\tablenotemark{1} 
     & 128$\times$224 & 0.05 & 0.30     &  1.0\phn &    3000 \\
sof6 & 128$\times$224 & 0.05 & 0.30     &  1.5\phn &    1800 \\
sof7 & 128$\times$224 & 0.05 & 0.30     &  2.0\phn &    1800 \\
sof8 & 128$\times$224 & 0.05 & 0.30     &  0.2\phn &    1800 \\
sof9 & 128$\times$224 & 0.05 & 0.30     &  0.4\phn &    1800 \\
sf10 & 128$\times$224 & 0.05 & 0.30     &  3.0\phn &    1800 \\
sf11 & 128$\times$224 & 0.05 & 0.30     &  4.0\phn &    1800 \\
Sof5 & 256$\times$448 & 0.05 & 0.30     &  1.0\phn &     250 \\
Sof7 & 256$\times$448 & 0.05 & 0.30     &  2.0\phn &    2400 \\
nosg\tablenotemark{2} 
     & 128$\times$224 & 0.05 & 0.30     &  1.0\phn &    1200 \\
Nosg\tablenotemark{2} 
     & 256$\times$448 & 0.05 & 0.30     &  1.0\phn &     500 \\
\enddata
\tablenotetext{1}{The labels {\it mas3} and {\it sof5} refer to the
same simulation, but referenced in two different series of
simulations.} 
\tablenotetext{2}{The simulations {\it nosg} and {\it Nosg} do not
include self gravity in the disk, but are otherwise identical to
simulations {\it mas3 (sof5)} and {\it Sof7}, respectively. } 

\end{deluxetable} 

\doublespace

\clearpage

\singlespace
\begin{deluxetable}{lcc}
\tablewidth{0pt}
\tablecaption{\label{tab:table-PW} Porter \& Woodward Table 1}
\tablehead{
\colhead{$C_a$}  & \colhead{$A_s$} & \colhead{$B_s$} 
}

\startdata
0.0008       &       343.71   &             4444.67 \\
0.008\phn    &    \phn35.36   &         \phn 496.68 \\
0.08\phn\phn & \phn\phn7.84   &         \phn 107.00 \\
0.16\phn\phn &    \phn10.84   &     \phn\phn  82.52 \\
0.32\phn\phn &    \phn20.37   &     \phn\phn  23.75 \\
0.56\phn\phn &    \phn34.70   & \phn\phn\phn   0.00 \\
\enddata
\end{deluxetable}
\doublespace

\clearpage

\singlespace

\clearpage

\clearpage

\clearpage

\clearpage

\clearpage

\clearpage

\clearpage

\clearpage

\clearpage

\clearpage

\clearpage

\clearpage

\end{document}